\documentclass[11pt,a4paper]{article}
\usepackage{jheppub}
\usepackage[utf8]{inputenc}
\usepackage{amsmath}
\usepackage{amsfonts}
\usepackage{amssymb}
\usepackage{graphicx}
\usepackage{float}
\usepackage{cancel}
\usepackage{extarrows}
\usepackage{caption}
\usepackage{subfig}
\usepackage{rotating}
\usepackage[toc,page]{appendix}
\usepackage{cleveref}

\graphicspath{{.}{images/}}

\usepackage{caption}
\makeatletter

\newcommand{\Rmnum}[1]{\expandafter\@slowromancap\romannumeral #1@}
\makeatother

\title{High scale boundary conditions with an additional complex singlet.}
\author{John McDowall,}
\author{David J. Miller}

\affiliation{SUPA, School of Physics and Astronomy, University of Glasgow, \\
						Glasgow, G12 8QQ, United Kingdom}

\emailAdd{j.mcdowall.1@research.gla.ac.uk}
\emailAdd{david.j.miller@glasgow.ac.uk}

\abstract{We investigate Planck scale boundary conditions on Higgs quartic interactions and their $\beta$-functions in the SM augmented by an additional complex scalar. We use renormalisation group running at two-loops, and include both theoretical and experimental constraints. We find that the boundary condition $\lambda=\beta_\lambda=0$ at the Planck scale is compatible with the current Higgs and top mass measurements, but requires additional scalars lighter than about $600\,$GeV.}

\begin{document}
\maketitle

%%%%%%%%%%%%%%%%%%%%%%%%%%%%%%%%%%%%%%%%%%%%%%%%%%%%%%%%%%%%%%%%%%%%%%%%%%%%%%%%%%%%%%%%%%%%%%%%%%%%%%%%%%%%%%%%
\section{Introduction}
\label{sec:introduction}
The Standard Model (SM) is an enormously successful descripton of the strong and electroweak interactions, and the discovery of the Higgs at the LHC \cite{Aad:2012tfa} is just one of a long list of experimental validations that it can boast. However, the combined ATLAS and CMS value of the Higgs mass \citep{Aad:2015zhl}, $m_h = 125.09 \pm 0.23\,$GeV, is in a phenomenologically tricky range for the most popular frameworks Beyond the Standard Model (BSM). Supersymmetric or composite Higgs models can provide a $125\,$GeV Higgs, but in general this requires either a large tuning of parameters or a non-minimal field content \citep{Buttazzo:2013uya,Craig:2013cxa,Ross:2017kjc}, leaving us in a situation where they are neither favoured nor excluded. Add to this the seemingly very SM-like nature of the Higgs \citep{Aad:2015zhl, Khachatryan:2014kca, Khachatryan:2014jba, Aad:2015gba} and we find searches for BSM physics in a difficult position. Many extensions of the Standard Model try to deal with its shortcomings by introducing new fields at around the TeV scale. However, these extensions must now evade increasingly strong experimental bounds on couplings and interactions.

\begin{figure}[htb]
  \centering
  \subfloat[]{\includegraphics[width=0.5\textwidth]{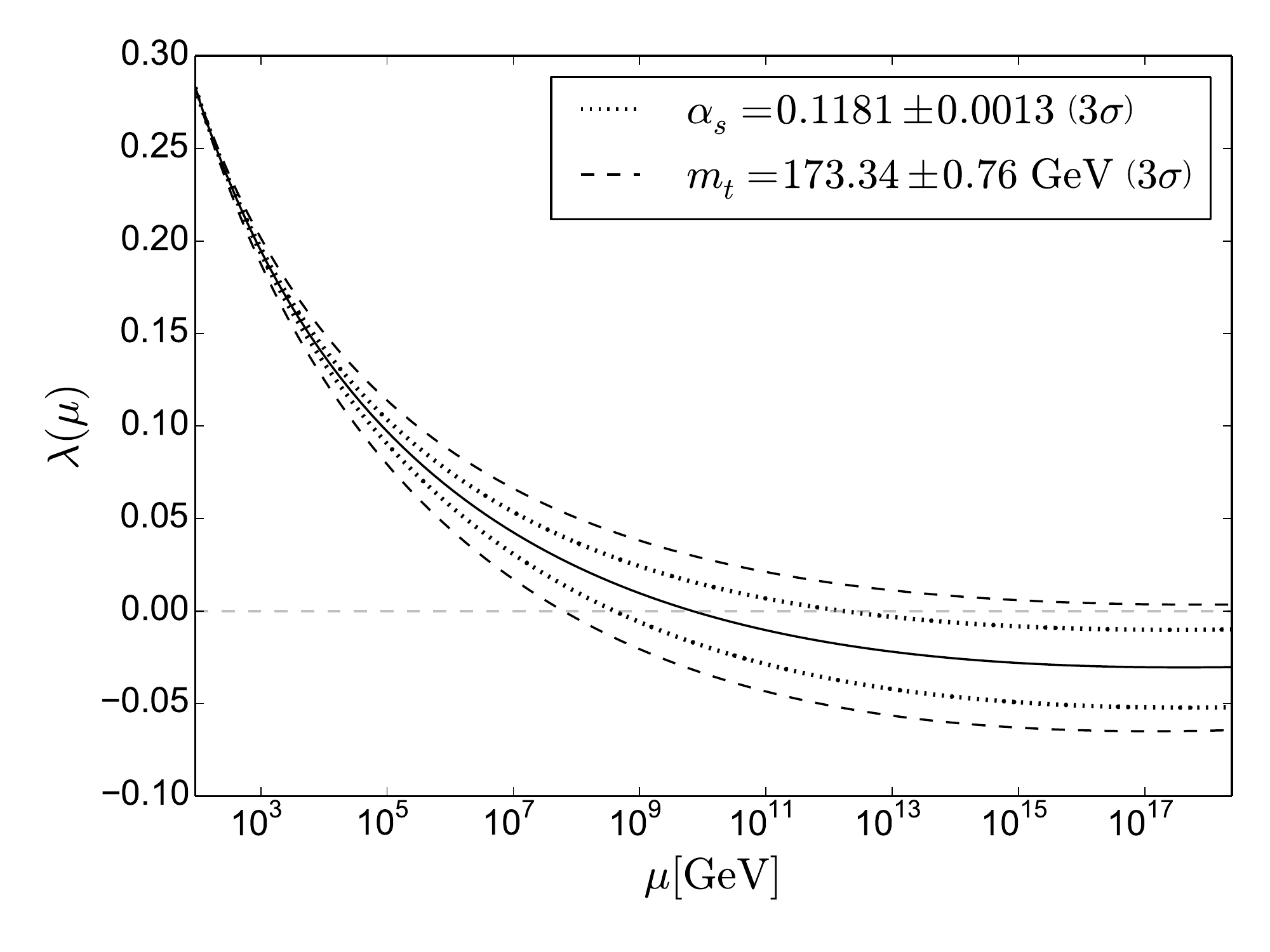} \label{fig:SM_lambda}}
  \subfloat[]{\includegraphics[width=0.5\textwidth]{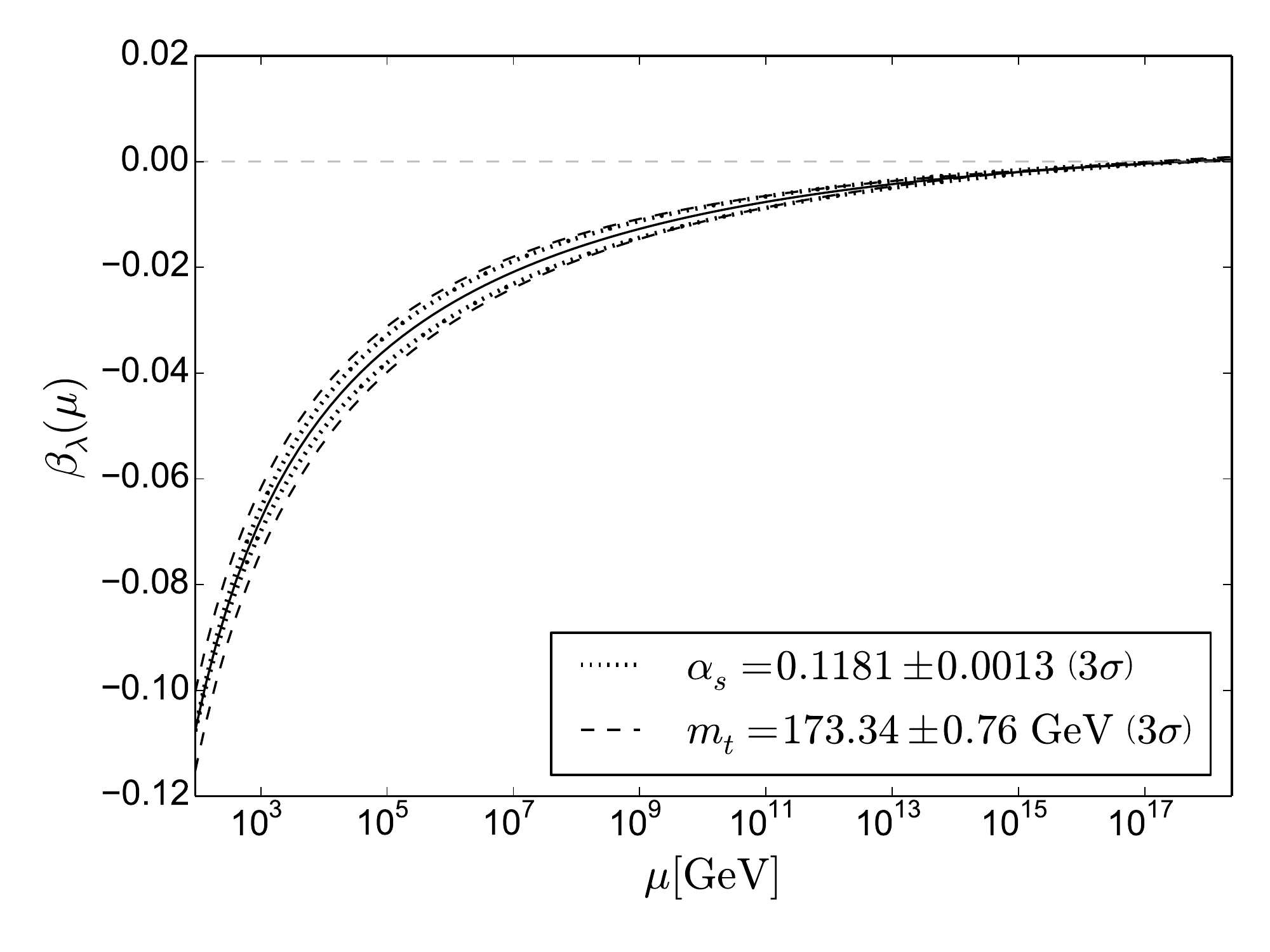} \label{fig:SM_betalambda}}
  \caption{Three-loop running of the SM Higgs quartic coupling $\lambda$ and its $\beta$ function with $3 \, \sigma$ uncertainties from the top pole mass $m_t$ (dashed) and the strong coupling constant $\alpha_s$ (dotted). Calculated  using SARAH $4.9.3$ \cite{Staub:2013tta} and FlexibleSUSY $1.6.1$ \cite{Athron:2014yba,Athron:2017fvs,Allanach:2001kg,Allanach:2013kza}.}
  \label{fig:SM_lambda_beta}
\end{figure}

The current lack of experimental evidence for BSM physics encourages exploration of whether the SM is a valid description of physics up to very high energies such as the Planck scale, $M_{\rm Pl}$. Figure \ref{fig:SM_lambda_beta}(a) shows that using the central values of the top pole mass $m_t$ and the strong coupling constant $\alpha_s \left( M_Z \right)$ in the calculation of the running results in the Higgs quartic coupling $\lambda$ turning negative at an energy scale $\mu \sim 10^{10}$ GeV. Requiring absolute stability of the vacuum puts a limit on the top mass \cite{Buttazzo:2013uya},
\begin{equation}
\label{eq:SM_topmassstabilitylimit}
m_t < 171.36 \pm 0.46 \text{ GeV},
\end{equation}
which is in tension with the current experimental value by about $2.6 \sigma$. It would seem that, assuming validity up to high scales, the SM has a metastable electroweak vacuum with a lifetime much longer than the age of the universe. If we insist upon a stable vacuum then we should consider metastability of the SM vacuum as an indication of some as yet unknown degrees of freedom that, when included in the RGE analysis, stabilises the potential.

An interesting feature of the running of the Higgs quartic coupling $\lambda$, shown in Figure \ref{fig:SM_lambda_beta}(a), is that at high scales $\lambda$ is very small. Not only that, but the running of $\lambda$ flattens out, i.e $\beta_{\lambda} \approx 0$ shown in Figure \ref{fig:SM_lambda_beta}(b). These properties of the Higgs quartic coupling has led to investigations into whether they are boundary conditions that are a consequence of some high scale dynamics at $M_{\rm Pl}$ \citep{Degrassi:2012ry,Holthausen:2011aa,Iacobellis:2016eof,Haba:2013lga,Eichhorn:2014qka,Khan:2014kba,Helmboldt:2016mpi}.

The main focus of this article is to consider whether a simple extension to the scalar sector of the SM can display behaviour that is in keeping with the existence of such high scale boundary conditions whilst also being compatible with theoretical and experimental constraints. To that end we add a complex singlet to the SM, which is the simplest extension that can provide a dark matter candidate as well as a first step towards an explanation for the matter anti-matter asymmetry of the universe via CP violation \citep{Darvishi:2016gvm}. The parameter space of this model allows for two distinct phases: one consisting of two scalars that mix and one dark matter candidate; and another where all three scalars mix freely. The more complicated potential and the altered running of the scalar quartic couplings also opens up the possibility of a stable vacuum. In this article we examine the parameter space of both phases, requiring points that are perturbative and stable up to the Planck scale, provide a SM Higgs candidate, and are not excluded via experimental collider and dark matter constraints. We further consider the scalar quartic couplings and their RGE evolution at high scales to determine if they display the behaviour indicative of high scale boundary conditions, and whether there are any measurable low energy consequences.

In section \ref{sec:SM_planck} we discuss in detail the connection between the high scale behaviour of the SM and the boundary conditions that may arise due to dynamics at that scale. In Section \ref{sec:complex_singlet} we detail the structure and phases of the Complex Singlet extension of the SM. In section \ref{sec:scan} we describe our methodology in performing the analysis of the parameter space, as well as the theoretical and experimental constraints that we apply. In sections \ref{sec:results_broken} and \ref{sec:results_DM} we discuss our results for the two phases of the model, and in section \ref{sec:conclusions} we make our conclusions.

%%%%%%%%%%%%%%%%%%%%%%%%%%%%%%%%%%%%%%%%%%%%%%%%%%%%%%%%%%%%%%%%%%%%%%%%%%%%%%%%%%%%%%%%%%%%%%%%%%%%%%%%%%%%%%%%
\section{Planck Scale Boundary Conditions and the Standard Model}
\label{sec:SM_planck}
We are primarily interested in the following boundary conditions in the Standard Model at $M_{\rm Pl}$,
\begin{eqnarray}
\lambda \left( M_{\rm Pl} \right) &= 0, \label{eq:SM_bc1} \\
\beta_{\lambda} \left( M_{\rm Pl} \right) &= 0. \label{eq:SM_bc2}
\end{eqnarray}
These conditions may arise for a number of reasons. For example, the \textit{Multiple Point Principle} (MPP) requires the effective potential have two degenerate vacua: the electroweak vacuum as well as a second vacuum at a high scale such as $M_{\rm Pl}$ \citep{Froggatt:1995rt}. Using the assumption $V_{\rm eff} \approx \frac{1}{4} \lambda \left( \mu \right) \phi^4$, where $\phi$ is the Higgs field, a second minimum at $M_{\rm Pl}$ requires both conditions \ref{eq:SM_bc1} and \ref{eq:SM_bc2}  be satisfied,
\begin{equation}
\label{eq:SM_MPP}
\frac{d V_{\rm eff}}{d \phi} \bigg\vert_{\phi=M_{\rm Pl}} = \lambda \left( \mu \right) \phi^3 + \frac{1}{4} \beta_{\lambda} \left( \mu \right) \phi^4 = 0.
\end{equation}
The MPP hypothesis gave an early prediction of the Higgs mass of $m_h = 135 \pm 9$ GeV, while a more recent calculation of the Higgs mass using a newer top-quark mass gave a prediction of $m_h = 129 \pm 1.5$ GeV \cite{Buttazzo:2013uya}. Although this is not compatible with our by now very accurate Higgs mass measurement, it is still rather remarkable. Subsequent investigations into this principle extend the SM with new fields to alter the running of $\lambda$ in the hope of satisfying the MPP conditions with a viable Higgs mass, for example \citep{Froggatt:2006zc, Kawana:2014zxa, Haba:2014sia, Hamada:2014xka, Haba:2016gqx, Nielsen:2017ows}.

Another high scale suggestion is to enforce \textit{Asymptotic Safety} in the SM, possibly using gravitational contributions to the RGEs at $M_{Pl}$ \citep{Shaposhnikov:2009pv,Wetterich:2011aa, Eichhorn:2017sok, Eichhorn:2017lry, Eichhorn:2017ylw}. In this scenario the running of the quartic coupling is altered at scales beyond some transition scale, e.g $M_{\rm Pl}$,
\begin{equation}
\mu \frac{d \lambda}{d \mu} = \beta_{\lambda}^{SM} + \beta_{\lambda}^{grav} = \beta_{\lambda}^{SM} + \frac{a}{8 \pi} \frac{\mu^2}{M_{Pl}^2} \lambda .
\label{eq:SM_AS}
\end{equation}
where $a$ encodes the nature of the gravitational contribution to the running. The coupling $\lambda$ may then reach an interacting UV fixed point, allowing the SM together with gravity to be valid up to arbitrarily high energies. The approach outlined in \citep{Shaposhnikov:2009pv} predicts a Higgs mass in the range $126\,{\rm GeV} < m_h < 174\,$GeV, with the intriguing lower limit arising when both boundary conditions \ref{eq:SM_bc1} and \ref{eq:SM_bc2} are met. However, it is not clear that the above gravitational contribution should really be used above $M_{\rm Pl}$, where effective quantum gravity no longer applies. There also remains a question over its gauge dependence.

Asymptotic safety requires an interacting UV fixed point, and should not be confused with \textit{Asymptotic Freedom}, where couplings are required to flow to zero at high scales. Total Asymptotic Freedom (TAF), where {\em all} the interaction couplings tend to zero, would also allow the model to be valid up to infinite energies. The fact that the hypercharge coupling of the SM eventually hits a Landau pole means that extensions which exhibit TAF need to embed the SM $U(1)$ into some larger group such as in a Pati-Salam model \citep{Giudice:2014tma, Einhorn:2017jbs, Holdom:2014hla, Gies:2015lia}. There have also been a number of recent investigations into Total Asymptotic Safety (TAS) in Ref.~\citep{Litim:2014uca, Litim:2015iea, Bond:2016dvk, Bond:2017wut, Sannino:2014lxa, Bajc:2016efj, Pelaggi:2017wzr, Pelaggi:2017abg}. Models that exhibit TAS may use just the SM gauge group and usually involve adding large numbers of scalars and vector-like fermions in a variety of representations.

Both the TAF and TAS scenarios are more involved than the simple extension we consider in this article. Here we consider boundary conditions generally, with an eye on their consequences at scales that can be directly probed by experiment, rather than focusing on the specific UV mechanism from which they arise.

\begin{figure}[!tbp]
\centering
\subfloat[]{\includegraphics[width=0.5\textwidth]{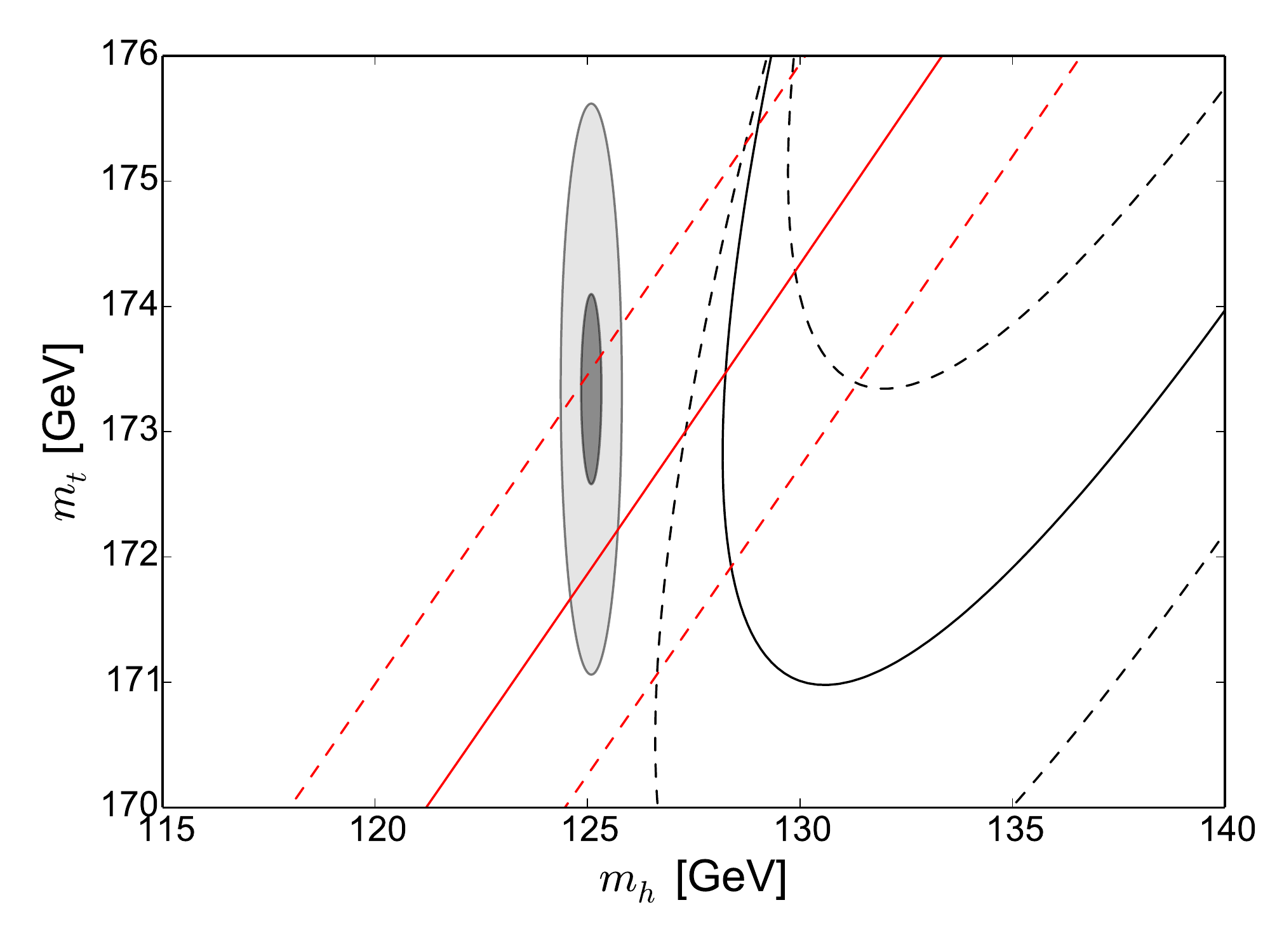} \label{fig:SM_lambdabeta}}
\subfloat[]{\includegraphics[width=0.5\textwidth]{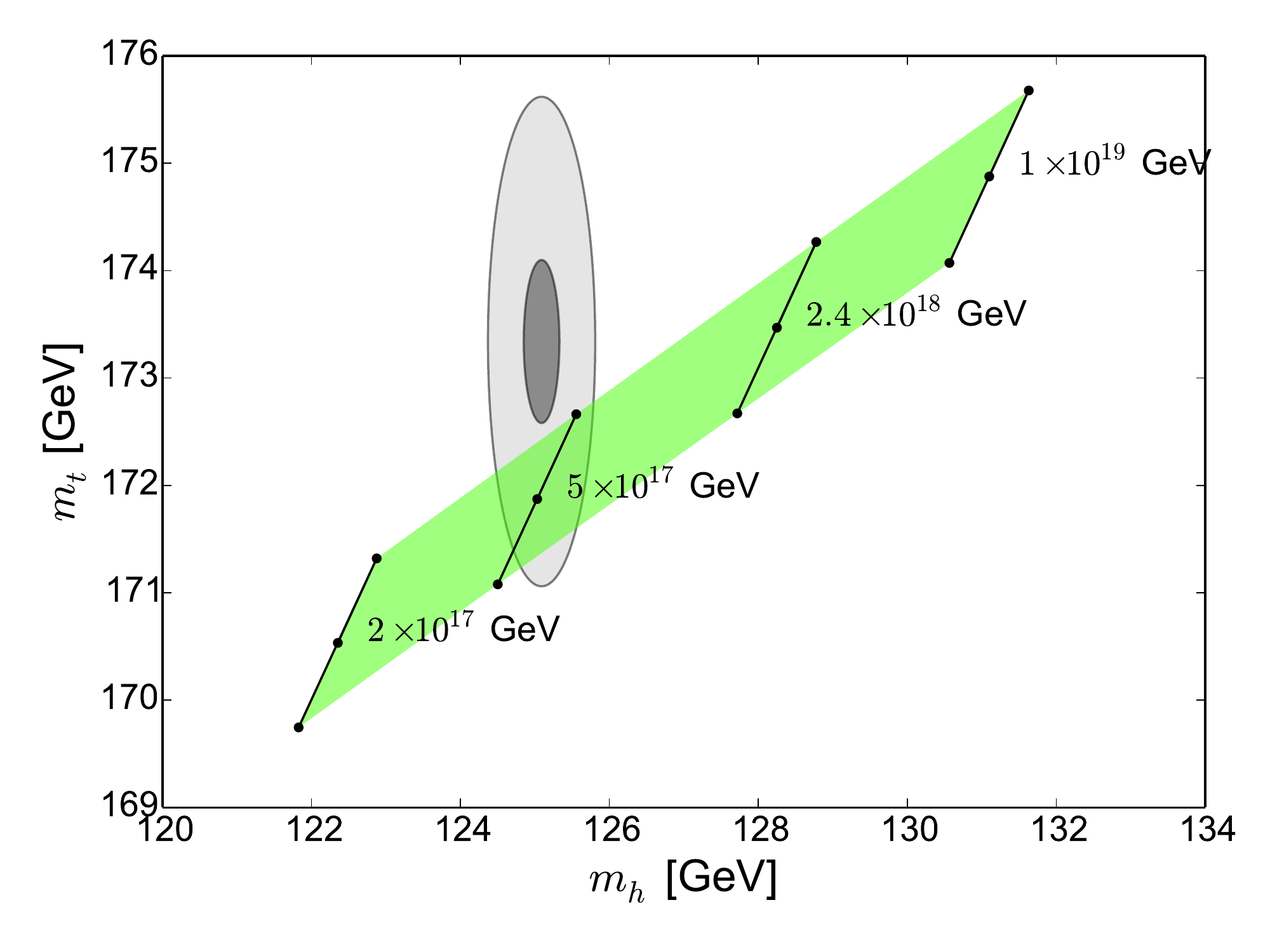} \label{fig:SM_scale}}
\caption{\textbf{(a)} $\lambda \left( M_{Pl} \right) = 0$ (red) and $\beta_{\lambda} \left( M_{Pl} \right) = 0$ (black) contours in the $m_h-m_t$ plane. The dashed lines show $3 \,\sigma$ variations in $\alpha_s \left( M_Z \right) = 0.1181 \pm 0.0013$. \textbf{(b)} Mass values that satisfy $\lambda(M_{\rm UV}) = \beta_{\lambda}(M_{\rm UV}) = 0$ at various UV scales $M_{UV}$. The green region corresponds to a $1\, \sigma$ uncertainty in $\alpha_s$. Ellipses show the experimentally allowed values of $m_t$ and $m_h$ with $1 \,\sigma$ (dark grey) and $3\,\sigma$ (light grey) uncertainty.}
\label{fig:SM_lambdabetalambda}
\end{figure}

Figure \ref{fig:SM_lambdabetalambda} shows contours corresponding to the boundary conditions \ref{eq:SM_bc1} and \ref{eq:SM_bc2} in the $m_h-m_t$ plane, indicating that a Higgs heavier than that experimentally observed is needed to require both conditions are satisfied, using the central values of $m_t$ and $\alpha_s$. These contours are calculated using three-loop SM RGEs; the Higgs mass is calculated to two-loop order, while the top mass additionally contains three-loop QCD corrections. This plot is in agreement with the similar plot in Ref.~\citep{Degrassi:2012ry}, but we use a different value of the uncertainty in the strong coupling constant $\alpha_s \left( M_Z \right) = 0.1181 \pm 0.0013$ to reflect more recent estimates \cite{PDG:2015}. We also use the reduced Planck scale $M_{\rm Pl} = 2.4 \times 10^{18}\,$GeV as our scale at which these boundary conditions are set. Figure \ref{fig:SM_lambdabetalambda}(a) shows that $\lambda \left( M_{\rm Pl} \right) = 0$ can be satisfied with an acceptable value of $m_h$ for a top mass $171\, {\rm GeV} \lesssim m_t \lesssim 174\,$GeV, and although the corresponding value of $\beta_{\lambda} \left( M_{\rm Pl} \right)$ is not zero, it is extremely small.

So far (and for the rest of this paper) we've required that these boundary conditions are satisfied at $M_{\rm Pl}$, but it is possible that the mechanisms by which these conditions come about become relevant at scales lower than $M_{\rm Pl}$. Figure \ref{fig:SM_lambdabetalambda}(b) shows points in the $m_h-m_t$ plane that satisfy both boundary conditions $\lambda = \beta_{\lambda} = 0$ simultaneously at different UV scales $M_{\rm UV}$. The green region corresponds to a $1\, \sigma$ uncertainty in $\alpha_s$. We see it is possible to obtain a Higgs mass that is within experimental limits by applying these boundary conditions at $M_{\rm UV} \approx 5 \times 10^{17}$ GeV. It's interesting to note that this is a scale of importance in string scenarios (see e.g \citep{Ginsparg:1987ee, Witten:1996mz}).

%%%%%%%%%%%%%%%%%%%%%%%%%%%%%%%%%%%%%%%%%%%%%%%%%%%%%%%%%%%%%%%%%%%%%%%%%%%%%%%%%%%%%%%%%%%%%%%%%%%%%%%%%%%%%%%%
\section{The Complex Singlet Extension of the Standard Model}
\label{sec:complex_singlet}
We investigate the phenomenologically viable regions of parameter space in the complex singlet extension of the SM to examine the effect of high scale boundary conditions. We include a complex scalar field $\mathbb{S} = S_1 + i S_2$ that is a singlet under the SM gauge group, and consider a scalar potential of the form
\begin{equation}
\label{eq:complex_potential}
V  = \frac{\mu^2}{2} H^{\dagger} H + \frac{\lambda}{4} \left( H^{\dagger} H \right)^2 + \frac{\delta}{2} \left( H^{\dagger} H \right) \left| \mathbb{S} \right|^2 + \frac{b_2}{2} \left| \mathbb{S} \right|^2 + \frac{d_2}{4} \left| \mathbb{S} \right|^4
+ \left( \frac{b_1}{4} \mathbb{S}^2 + a_1 \mathbb{S} + c.c \right)
\end{equation}
This model was first discussed in Ref.~\cite{Barger:2008jx}, where a $U(1)$ symmetry was used to simplify the model and remove complex phases. This $U(1)$ is softly broken by the term $\propto b_1$ and $a_1$. For other recent investigations into the phenomenology of this model see, for example, Refs. \citep{Barger:2008jx, Barger:2010yn, Gonderinger:2012rd, Costa:2014qga, Coimbra:2013qq, Robens:2015gla, Muhlleitner:2017dkd}

Our implementation of this model in FlexibleSUSY is more straightforward if we consider the equivalent theory in terms of two real scalars. Therefore we make use of the parameter definitions
\begin{equation}
	b_{\pm} = \frac{1}{2} \left (b_2 \pm b_1 \right),
\end{equation}
which function as (squared) masses for the two real scalar fields. During electroweak symmetry breaking, the complex singlet field acquires a non-zero vacuum expectation value (vev) for its real, and possibly imaginary, part. If both real and imaginary parts acquire non-zero vevs,
\begin{equation}
\label{eq:complex_vevs}
\mathbb{S} = \frac{1}{\sqrt{2}} \left[ v_{s_1} + s_1 + i \left( v_{s_2} + s_2 \right) \right],
\end{equation}
we call this the {\em broken phase}, following the nomenclature of Ref.~\cite{Costa:2014qga}. Therefore, in addition to the bilinear terms $\mu^2$ and $b_{\pm}$ which are fixed via the electrowak vacuum minimisation conditions, the model is described by
\begin{equation}
\label{eq:complex_inputs_broken}
\lambda, \quad d_2, \quad \delta, \quad v_{s_1}, \quad v_{s_2}, \quad a_1.
\end{equation}
In this phase, all three scalar field fluctuations $h$, $s_1$ and $s_2$ mix.

In contrast, if the vev of the imaginary part remains zero, the second electroweak vacuum minimisation condition (for $S_2$) is trivial and $b_-$ becomes a free parameter. In this case the input parameters are
\begin{equation}
\label{eq:complex_inputs_DM}
\lambda, \quad d_2, \quad \delta, \quad v_{s_1}, \quad b_{-}, \quad a_1.
\end{equation}
Now we find ourselves in the {\em dark matter} (DM) {\em phase}, where mixing is allowed only between $h$ and the real part of the complex singlet field $s_1$. The imaginary part $s_2$ does not mix and is a dark matter candidate kept stable by the symmetry $S_2 \to - S_2$.

In this work we describe each point in parameter space using the potential couplings, but we note that in some circumstances it may be more useful to use the Higgs masses, vevs and mixing angles as inputs. We also allow the top pole mass $m_t$ and the strong coupling constant $\alpha_s \left( M_Z \right)$ to vary as input parameters by $\pm 3 \sigma$ of their central values to take into account their contribution to the uncertainty in our results.

To investigate the RGE evolution of the scalar quartic couplings we use $\beta$ functions calculated to two-loops by SARAH. The gauge coupling $\beta$ functions in this model are identical to those in the SM. The running of the Yukawa couplings deviates only very slightly from the SM, and has no significant impact on our study, but is included for completeness. These $\beta$ functions are detailed in Appendix \ref{Appendix:RGEs}.

\begin{table}[!tbp]
\centering
\begin{tabular}{|l|l|}
\hline
\multicolumn{2}{|c|}{Broken Phase Input} \\ \hline
$\lambda \left( M_Z \right)$ & $0 - 0.5$         \\ \hline
$d_2 \left(M_{Pl} \right)$ & $0 - 0.5$         \\ \hline
$\delta \left( M_{Pl} \right)$ & $0 - 0.5$        \\ \hline
$v_{s_1}$ & $0 - 2000$ GeV         \\ \hline
$v_{s_2}$ & $0 - 2000$ GeV         \\ \hline
$a_1$ &   $-\left(10^{8} - 0 \right)$ $\mathrm{GeV}^3$ \\ \hline
\end{tabular} \qquad
\begin{tabular}{|l|l|}
\hline
\multicolumn{2}{|c|}{DM Phase Input} \\ \hline
$\lambda \left( M_Z \right)$ & $0 - 0.5$         \\ \hline
$d_2 \left(M_{Pl} \right)$ & $0 - 0.5$         \\ \hline
$\delta \left( M_{Pl} \right)$ & $0 - 0.5$        \\ \hline
$v_{s_1}$ & $0 - 2000$ GeV         \\ \hline
$b_{-}$ & $0 - 10^{5}$ $\mathrm{GeV}^2$         \\ \hline
$a_1$ &   $-\left(10^{8} - 0\right)$ $\mathrm{GeV}^3$ \\ \hline
\end{tabular}
\caption{Input parameter ranges for the numerical analysis of the \textbf{(left)} broken and \textbf{(right)} DM phases.}
\label{tab:complex_parameter_ranges}
\end{table}

\section{Numerical Analysis and Constraints}
\label{sec:scan}
We are interested here in the effect of boundary conditions on some or all of the model's quartic couplings at the Planck scale. In particular, these are:

\begin{eqnarray}
\label{eq:complex_bc1}
\lambda, \delta,  d_2 &= 0, \\
\label{eq:complex_bc2}
\beta_{\lambda}, \beta_{\delta}, \beta_{d_2} &= 0.
\end{eqnarray}
%
%We investigate the range of input parameters detailed in Table \ref{tab:parameter_ranges},
We fix values of $d_2$ and $\delta$ at the Planck scale, $\lambda$ at $M_Z$, as well as low scale values of $v_{s_1}$, $v_{s_2}$ (in the broken phase), $b_-$ (in the DM phase) and $a_1$,
and calculate the resulting mass spectrum, applying phenomenological and experimental constraints to each point. We use the Mathematica package SARAH $4.9.3$ \cite{Staub:2013tta} to calculate the $\beta$ functions at two loops for all of the model parameters. SARAH also calculates all of the mass matrices, tadpole equations, vertices and loop corrections required by spectrum generators to provide the mass spectrum. We use FlexibleSUSY $1.6.1$ \cite{Athron:2014yba,Athron:2017fvs,Allanach:2001kg,Allanach:2013kza} to build a spectrum generator using the SARAH output, and calculate the mass spectrum. This requires repeated running between $M_Z$ and the reduced Planck scale $M_{Pl}$ to obtain convergence of low and high energy parameters. The input parameter ranges for our scan of poth the broken and DM phases are detailed in Table \ref{tab:complex_parameter_ranges}

Valid scenarios must result in a vacuum that is bounded from below up to $M_{Pl}$. To that end we require the parameters satisfy three stability conditions at all scales \citep{Gonderinger:2012rd}
\begin{eqnarray}
\label{eq:complex_vacuumstability1}
\lambda &\geq& 0, \\ \label{eq:complex_vacuumstability2}
d_2 &\geq& 0, \\ \label{eq:complex_vacuumstability3}
\delta + \sqrt{\lambda d_2} &\geq& 0 .
\end{eqnarray}
We also require all of the dimensionless couplings of our model remain perturbative up to $M_{Pl}$. Specifically, for the Higgs quartic couplings this implies, at all scales,
\begin{equation}
\label{eq:complex_perturbative}
\lambda, \delta, d_2 < \sqrt{4 \pi}.
\end{equation}
We further check the stability of the vacuum using Vevacious \citep{Camargo-Molina:2013qva} which minimises the one-loop effective potential for each scenario and checks if the EWSB minimum is the global minimum. We accept points that provide a stable vacuum up to $M_{Pl}$. Finally, we insist the mass spectrum of valid points must contain an SM-like Higgs candidate with mass $124.7 \, {\rm GeV} \leq m_{h_{SM}} \leq 127.1 \,$GeV. This is a wider range than the experimental uncertainty in order to incorporate an estimate of the theoretical error.

Perturbativity, vacuum stability and the existence of an SM Higgs candidate are strong theoretical constraints on the parameter space of this model, invalidating the vast majority of the points investigated by our numerical analysis. However, since our goal is to examine scenarios that are phenomenologically compatible with current experiment, we must also apply constraints from the LHC Run-I, LEP and the Tevatron. We use HiggsBounds $4.3.1$ \citep{Bechtle:2013wla} and HiggsSignals $1.4.0$ \citep{Bechtle:2013xfa} to apply these constraints. The required inputs for both of these are,
\begin{equation}
\label{eq:complex_HBHS_inputs}
m_{h_i}, \quad \Gamma_{\mathrm{total}} \left( h_i \right), \quad \mathrm{BR} \left( h_i \rightarrow \mathrm{SM} \right), \quad  \mathrm{BR} \left( h_i \rightarrow h_j h_k \right), \quad \frac{\sigma \left( h_i \right)}{\sigma_{\mathrm{SM}} \left( h_{i} \right)},
\end{equation}
i.e.\ the scalar masses, their total decay widths, their branching ratios to SM fermions, gauge bosons and the other scalars, and their production cross-sections for all production modes, normalised to the SM production cross-sections evaluated at $m_{h_i}$.

We use sHDECAY \citep{Costa:2015llh, Djouadi:1997yw, Butterworth:2010ym} to calculate the branching ratios and total decay widths for each of our scenarios. The couplings of the scalars $h_i$ to the SM particles are suppressed with respect to the SM by the entry $R_{i1}$ of the Higgs mixing matrix (see Appendix~\ref{Appendix:mixing}). Therefore the signal strength is reduced to
\begin{equation}
\mu_i = R^2_{i1} \frac{R^2_{i1} \Gamma \left( h_{SM} \rightarrow \mathrm{SM} \right)}{R^2_{i1} \Gamma \left( h_{SM} \rightarrow \mathrm{SM} \right) + \sum \Gamma \left( h_{i} \rightarrow h_j h_k \right)},
\label{eq:complex_HBHS_signal_strength}
\end{equation}
which reduces further to $R^2_{i1}$ if decays to new scalars are forbidden. The decay widths to scalars are \citep{Coimbra:2013qq},
\begin{eqnarray}
\label{eq:complex_HBHS_decay_new_scalars}
\Gamma \left(h_i \rightarrow h_j h_j \right) &=& \frac{g^2_{ijj}}{32 \pi m_i} \sqrt{1 - \frac{4 m^2_j}{m^2_i}} \\
\Gamma \left(h_i \rightarrow h_j h_k \right) &=& \frac{g^2_{ijk}}{16 \pi m_i} \sqrt{1 - \frac{(m_j + m_k)^2}{m^2_i}} \sqrt{1 - \frac{(m_j - m_k)^2}{m^2_i}}
\end{eqnarray}
where $g_{ijj}$ and $g_{ijk}$ are the couplings between the scalars.

Inputing these into HiggsBounds provides $95 \%$ exclusion limits arising from the decay of new scalar states. Note, however, that the scalar-to-scalar decays in HiggsBounds version 4.3.1 includes only exclusion limits arising from decays of scalars to {\em identical} scalars $h_i \rightarrow h_j h_j$ \footnote{A beta version of HiggsBounds $5$ is now available  which does support decays to different scalars, however there are currently no experimental results available for such signatures. We also note that HiggsBounds 4.3.1 and HiggsSignals 1.4.0 only include LHC Run-I data.} (For recent work on di-Higgs production with different masses see \citep{Dawson:2017jja}.) HiggsSignals uses the same input as HiggsBounds to calculate a $\chi^2$ comparison to the observed SM Higgs production, which we also convert to a $95 \%$ exclusion.

The DM phase of the model must also conform to bounds arising from the DM relic density, which we calculate using micrOMEGAS \citep{Belanger:2014vza} and compare to the combined WMAP \citep{Hinshaw:2012aka} and Planck \citep{Ade:2015xua} experimental results,
\begin{equation}
\label{eq:complex_DM_exp_result}
\Omega h^2 = 0.1199 \pm 0.0027.
\end{equation}
The DM annihilation processes that contribute to the relic density include \textit{s}-channel annihilations into fermions, $W^+ W^-$ and $ZZ$, as well as to the SM higgs via the \textit{s}, \textit{t}, and \textit{u}-channels. We exclude points with results greater than $\Omega h^2 + 3 \sigma$, allowing for the possibility that the scalar is not the only contribution to the dark matter relic density but ensuring that our DM candidate does not overclose the universe.

Additionally we impose constraints from DM direct detection experiments that place limits on the spin independent scattering cross section $\sigma^{SI}$ of weakly interacting massive particles (WIMPs) on nucleons. The strongest constraints on WIMP DM from direct detection currently come from the LUX experiment \citep{Akerib:2016vxi} and are dependent on the mass of the DM candidate. The cross-section for a WIMP dark matter candidate on a proton can be calculated using \citep{Barger:2010yn},

\begin{equation}
\sigma^{SI} = \frac{m_p^4}{2 \pi \left( m_p + m_{h_{DM}} \right)^2} \left( \frac{\delta \left( b_1 - m_{h_{DM}}^2 \right)}{2 m_{h_1}^2 m_{h_2}^2} \right)^2 \left( \sum_{i=u,d,s} f_{\pi} + \frac{2}{27} \left( 3 f_G \right) \right)^2,
\label{eq:complex_DM_DD_sigmaSI}
\end{equation}
where $m_p$ is the proton mass, $m_{h_{DM}}$ is the DM candidate mass, $m_{h_i}$ are the remaining two Higgs masses, and $f_{\pi}, f_G$ are proton matrix elements \citep{Ellis:2000ds}. We use micrOMEGAS to calculate $\sigma^{SI}$ for each scenario and exclude those that result in a $\sigma^{SI}$ larger than the relevant limit from the 2016 LUX data.

%%%%%%%%%%%%%%%%%%%%%%%%%%%%%%%%%%%%%%%%%%%%%%%%%%%%%%%%%%%%%%%%%%%%%%%%%%%%%%%%%%%%%%%%%%%%%%%%
\section{The Broken Phase}
\label{sec:results_broken}
We now present results of our numerical analyses of the broken phase, in which all three neutral scalars mix, applying the theoretical and experimental constraints described in section \ref{sec:scan}. In this phase we call the SM-like Higgs $m_{h_{SM}}$, whilst the remaining two scalars are identified as $m_{h_{\rm Light}}$ and $m_{h_{\rm Heavy}}$, with $m_{h_{\rm Light}} < m_{h_{\rm Heavy}}$. (Note that $h_{\rm Light}$ may still be heavier than the SM-like Higgs, and correspondingly $h_{\rm Heavy}$ may be lighter.)

\begin{figure}[tbh]
  \centering
  \subfloat[]{\includegraphics[width=0.5\textwidth]{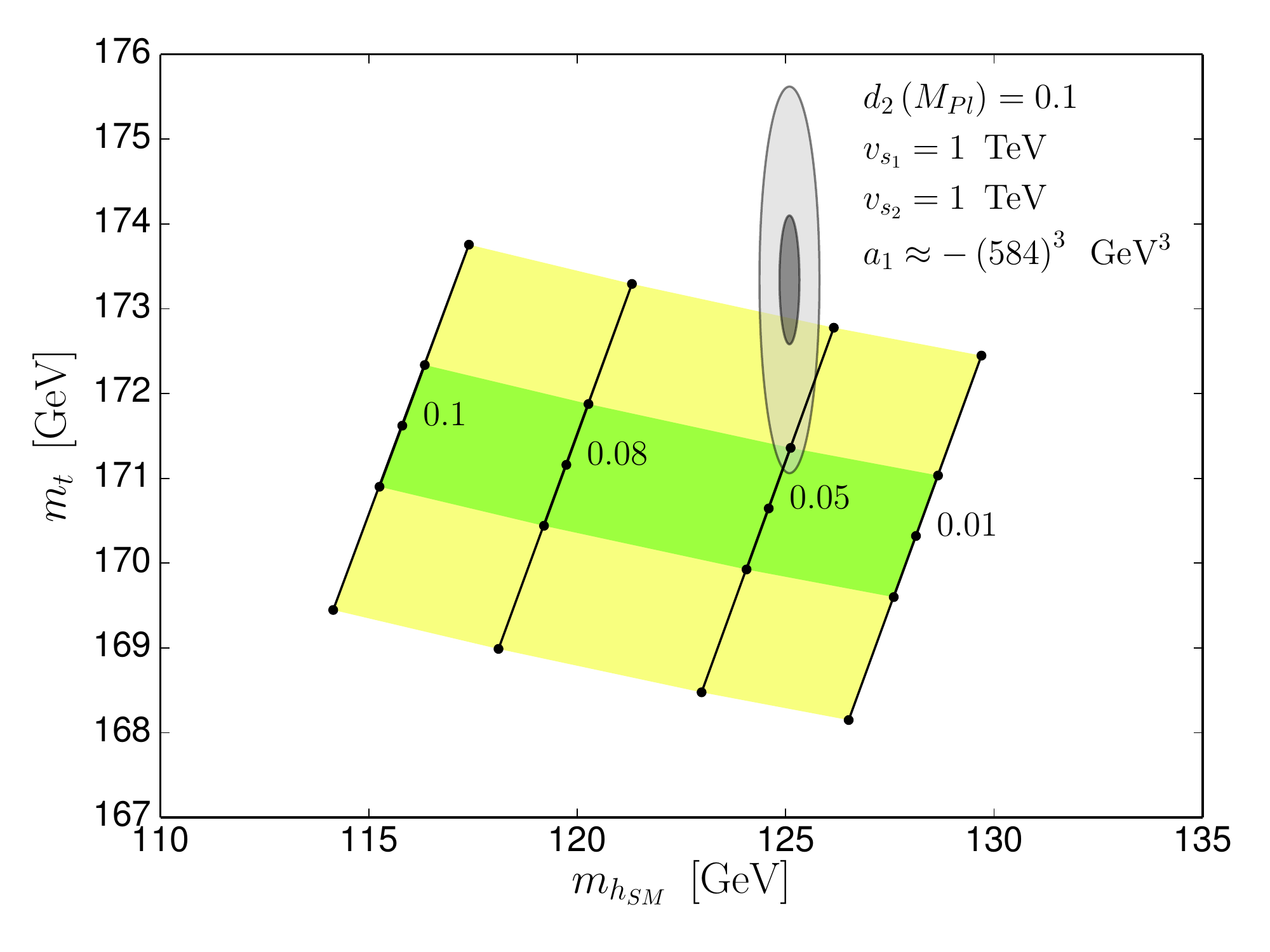}\label{fig:complex_d2_delta_int_1tev}}
  \hfill
  \subfloat[]{\includegraphics[width=0.5\textwidth]{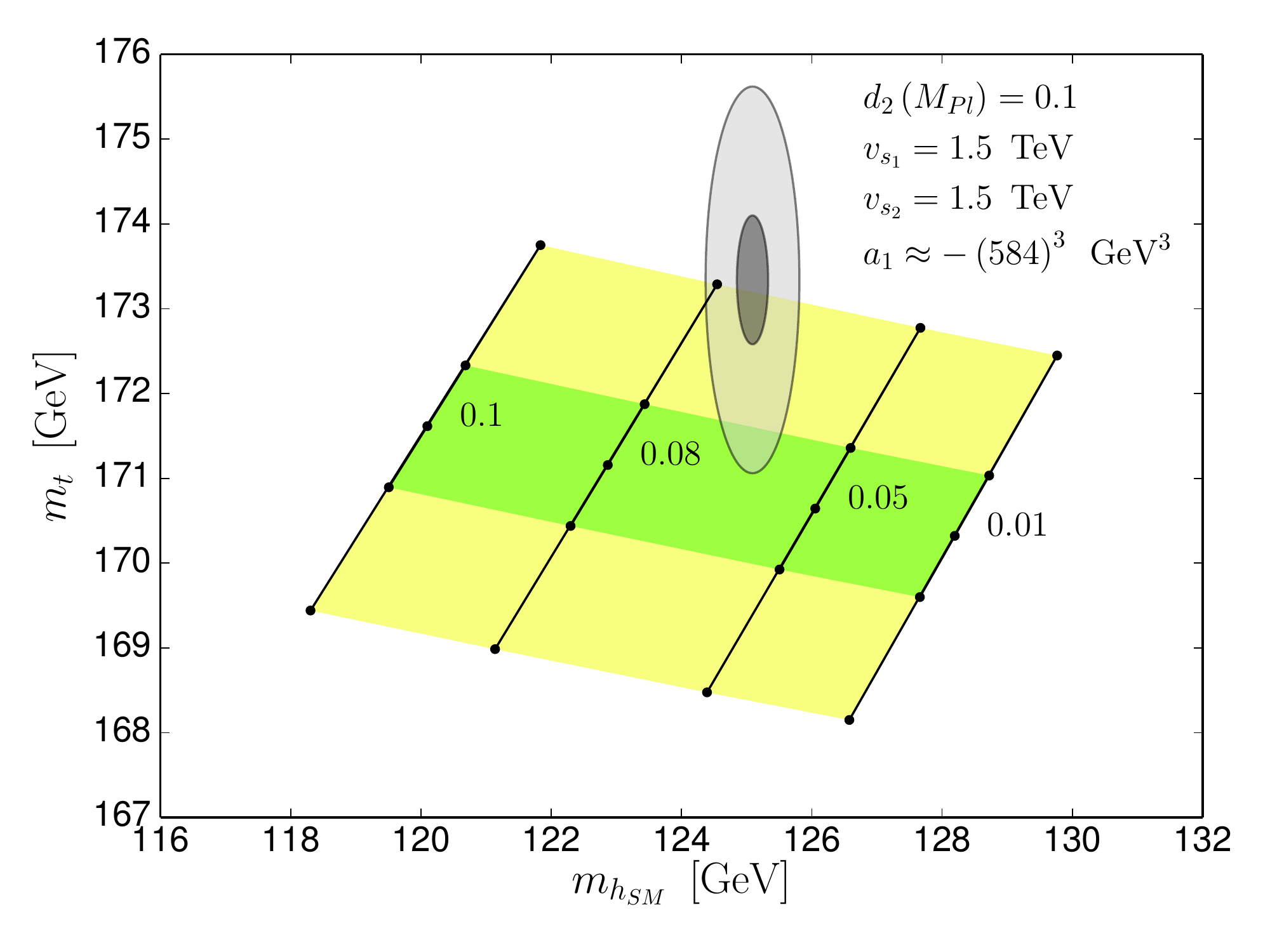}\label{fig:complex_d2_delta_int_1_5tev}}
  \hfill
    \subfloat[]{\includegraphics[width=0.5\textwidth]{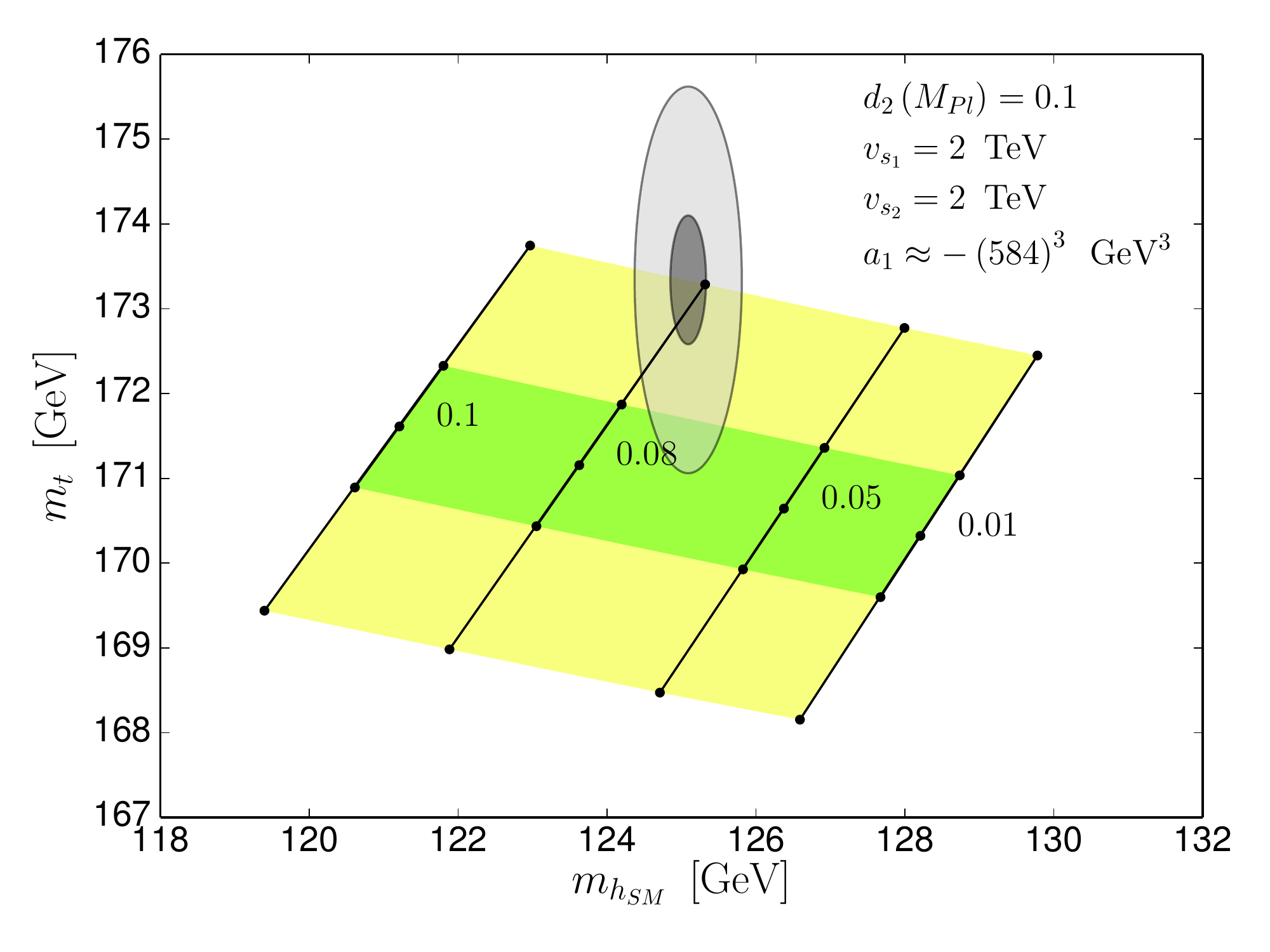}\label{fig:complex_d2_delta_int_2tev}}
  \hfill
  \subfloat[]{\includegraphics[width=0.5\textwidth]{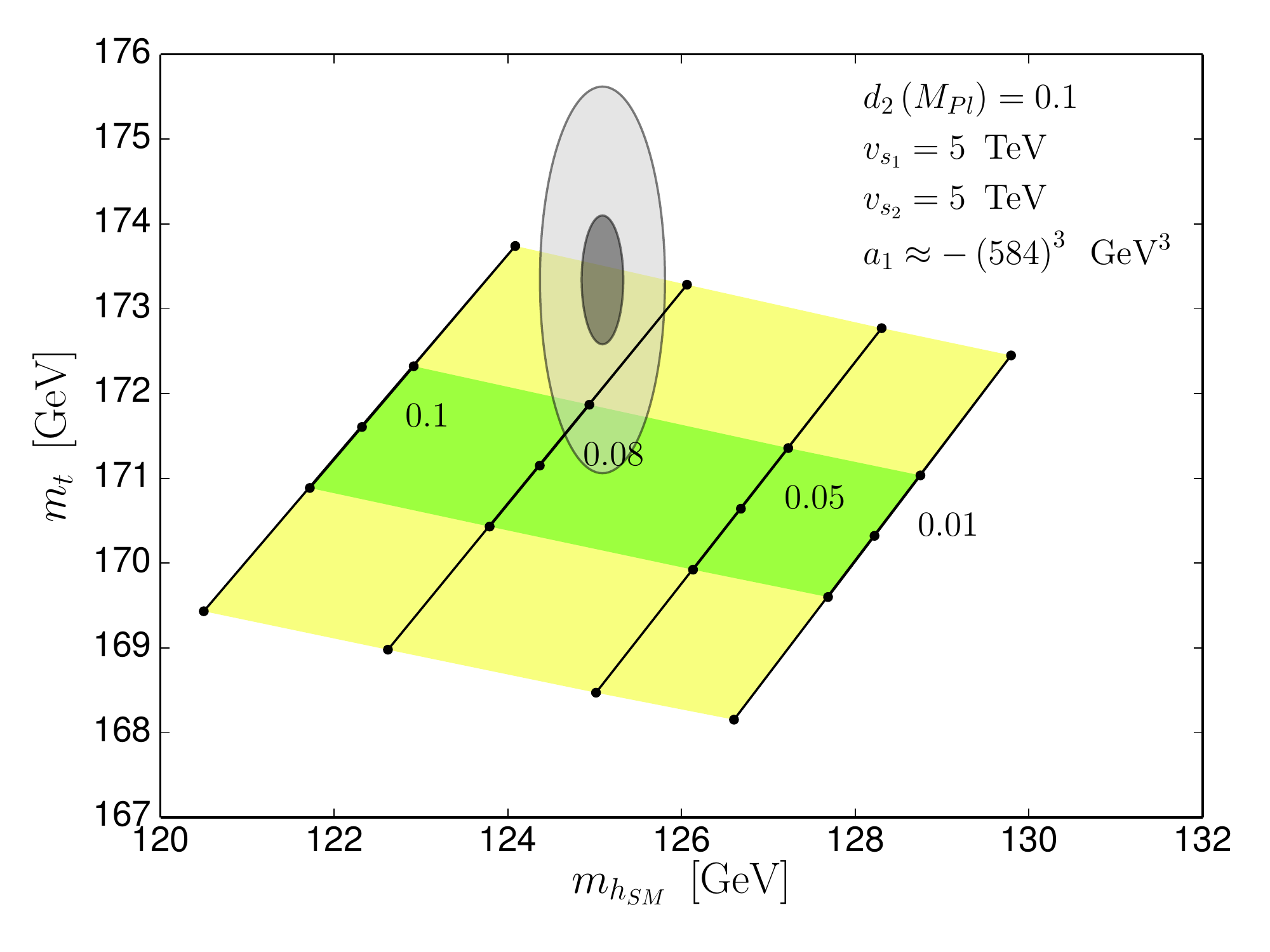}\label{fig:complex_d2_delta_int_5tev}}
\caption{Points in the $m_{h_{SM}}-m_t$ plane that satisfy $\lambda \left( M_{Pl} \right) = \beta_{\lambda} \left( M_{Pl} \right) = 0$ for $\delta \left( M_{Pl} \right) = 0.1, 0.08, 0.05, 0.01$ with \textbf{(a)} $v_{s_1} = v_{s_2} = 1$ TeV \textbf{(b)} $1.5$ TeV \textbf{(c)} $2$ TeV and \textbf{(d)} $5$ TeV. The green (yellow) region corresponds to $\pm 1 \left( 3 \right) \sigma $ uncertainty in $\alpha_s \left( M_Z \right) = 0.1181 \pm 0.0013$, whilst the ellipses show the experimentally allowed values of $m_t$ and $m_{h_{SM}}$ at $1 \,\sigma$ (dark grey) and $3 \,\sigma$ (light grey) uncertainty.}
\label{fig:complex_d2_delta_int}
\end{figure}

In figure \ref{fig:complex_d2_delta_int} we see $1 \,\sigma$ (green) and $3 \,\sigma$ (yellow) regions in the $m_{h_{SM}}-m_t$ plane of the broken phase that satisfy both boundary conditions $\lambda \left( M_{Pl} \right) = \beta_{\lambda} \left( M_{Pl} \right) = 0$ for different values of $v_{s_1}$ and $v_{s_2}$. The soft bilinear term $a_1$ and the complex singlet quartic coupling $d_2$ are kept fixed, whilst each line corresponds to a different high scale value of the Higgs portal coupling $\delta$. We see it is possible to satisfy the experimental constraints on the masses of both the top and the SM-like Higgs whilst meeting both high scale boundary conditions.

\begin{figure}[tbh]
  \centering
  \subfloat[]{\includegraphics[width=0.5\textwidth]{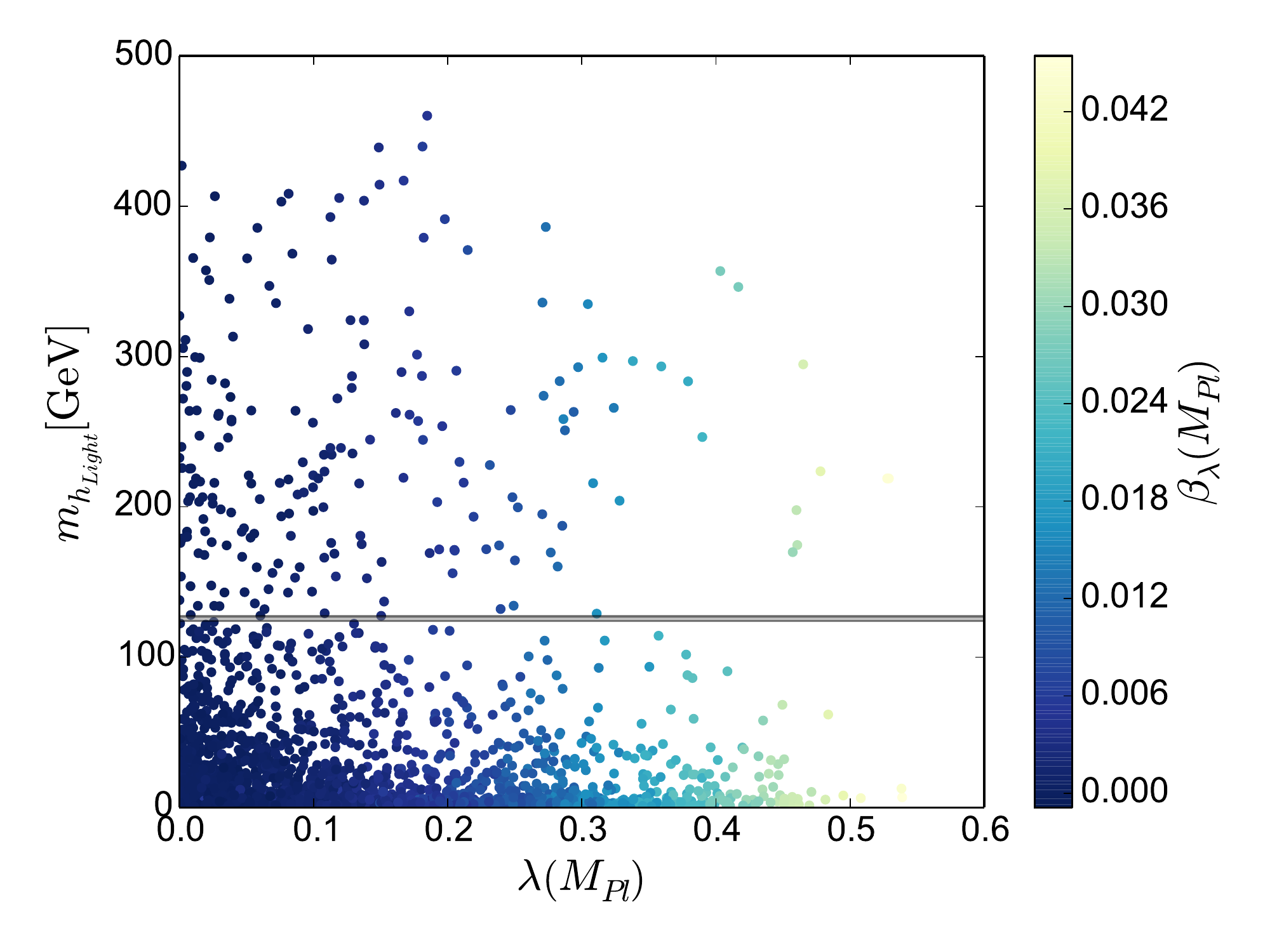}\label{fig:complex_lmplanck_lighthiggs_betal}}
  \hfill
  \subfloat[]{\includegraphics[width=0.5\textwidth]{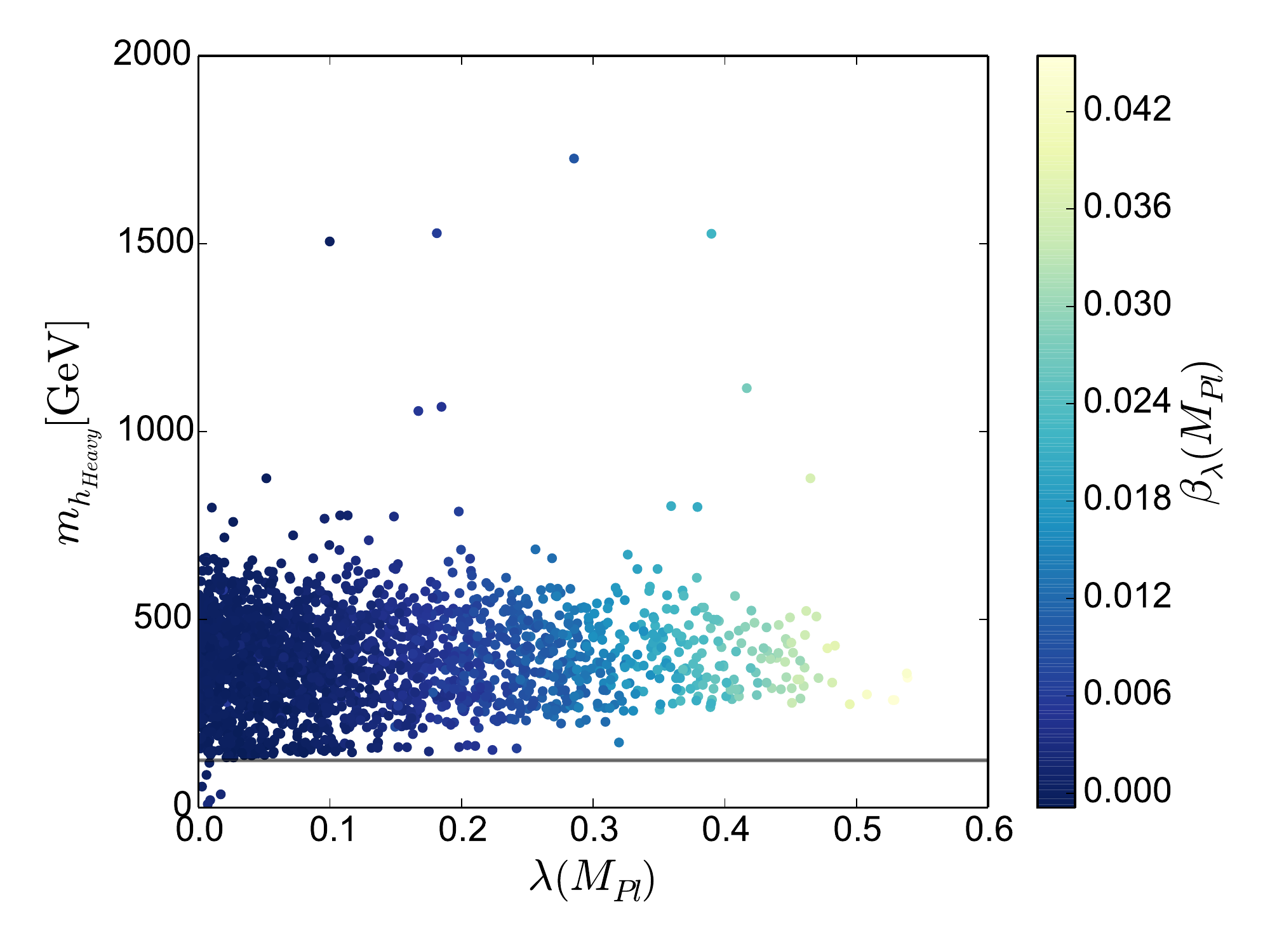}\label{fig:complex_lmplanck_heavyhiggs_betal}}
\caption{Higgs quartic coupling $\lambda \left( M_{\rm Pl} \right)$ and $\beta_{\lambda} \left( M_{\rm Pl} \right)$ compared to the light additional Higgs mass $m_{h_{\rm Light}}$ or the heavy additional Higgs mass $m_{h_{\rm Heavy}}$. Parameter points pass the theoretical and experimental constraints of Section~\ref{sec:scan}. The grey band shows the SM Higgs mass range. Only points with $\beta_\lambda(M_{\rm Pl}), \beta_\delta(M_{\rm Pl}), \beta_{d_2}(M_{\rm Pl}) \leq 0.05$ are shown. }
\label{fig:complex_lmplanck_betal}
\end{figure}

This compatibility is possible for a relatively large range of extra scalar masses. In Figure \ref{fig:complex_lmplanck_betal} we show light (a) and heavy (b) scalar masses resulting from a scan of parameters and their corresponding high scale values of $\lambda$ and $\beta_\lambda$, once the theoretical and experimental constraints have been applied. We allowed $d_2$ and $\delta$ (at $M_Z$) to vary between $0$--$0.5$, the vevs $v_{s_1}$ and $v_{s_2}$ between $0$-$2\,$TeV and $a_1$ between $-(464 \, {\rm GeV})^3$ and zero. $\lambda (M_Z)$ takes values between $0.2$ and $0.43$, potentially differing from the SM central value due to variation in the top mass, $\alpha_s$, and the new states. Here we are interested in the points with $\lambda = \beta_\lambda=0$ at the Planck scale, so those in dark blue to the left of the plots. For clarity of the plots we restrict them to only show scenarios for which $\beta_\lambda(M_{\rm Pl}), \beta_\delta(M_{\rm Pl}), \beta_{d_2}(M_{\rm Pl}) \leq 0.05$, but note that very few points exceed this value.

Before proceeding we should clarify what we really mean by a parameter, such as $\lambda$ or $\beta_\lambda$, being zero at $M_{\rm Pl}$. At first glance one might expect that we should set these parameters to be exactly zero at the Planck scale.  However, we must acknowledge that the relation between these parameters and our calculated low scale masses is necessarily perturbative. Therefore we should not restrict ourselves to keeping these parameters {\em exactly} zero but allow {\em small} values consistent with our uncertainty.

To provide an estimate on the uncertainty in our RGE evolution, we consider the difference between high scale parameters derived from the one-loop and two-loop Higgs quartic coupling RGEs (fixed at the low scale) and consider our parameters to be ``zero'' if they are smaller than this amount. For $\lambda$ this allows relatively large values of up to $0.067$, while for $\beta_\lambda$ we have a much tighter constraint of
\begin{equation}
	\beta_\lambda \lesssim 0.00005,
	\label{eq:betal_cutoff}
\end{equation}
so only the very darkest points of Figure \ref{fig:complex_lmplanck_betal} satisfy $\beta_\lambda=0$. One should not confuse this allowance with the uncertainty in $\lambda$ or $\beta_\lambda$ due to the top quark mass or $\alpha_s$, which are already taken into account when applying low energy constraints.

\begin{figure}[tbh]
  \centering
  \includegraphics[width=0.5\textwidth]{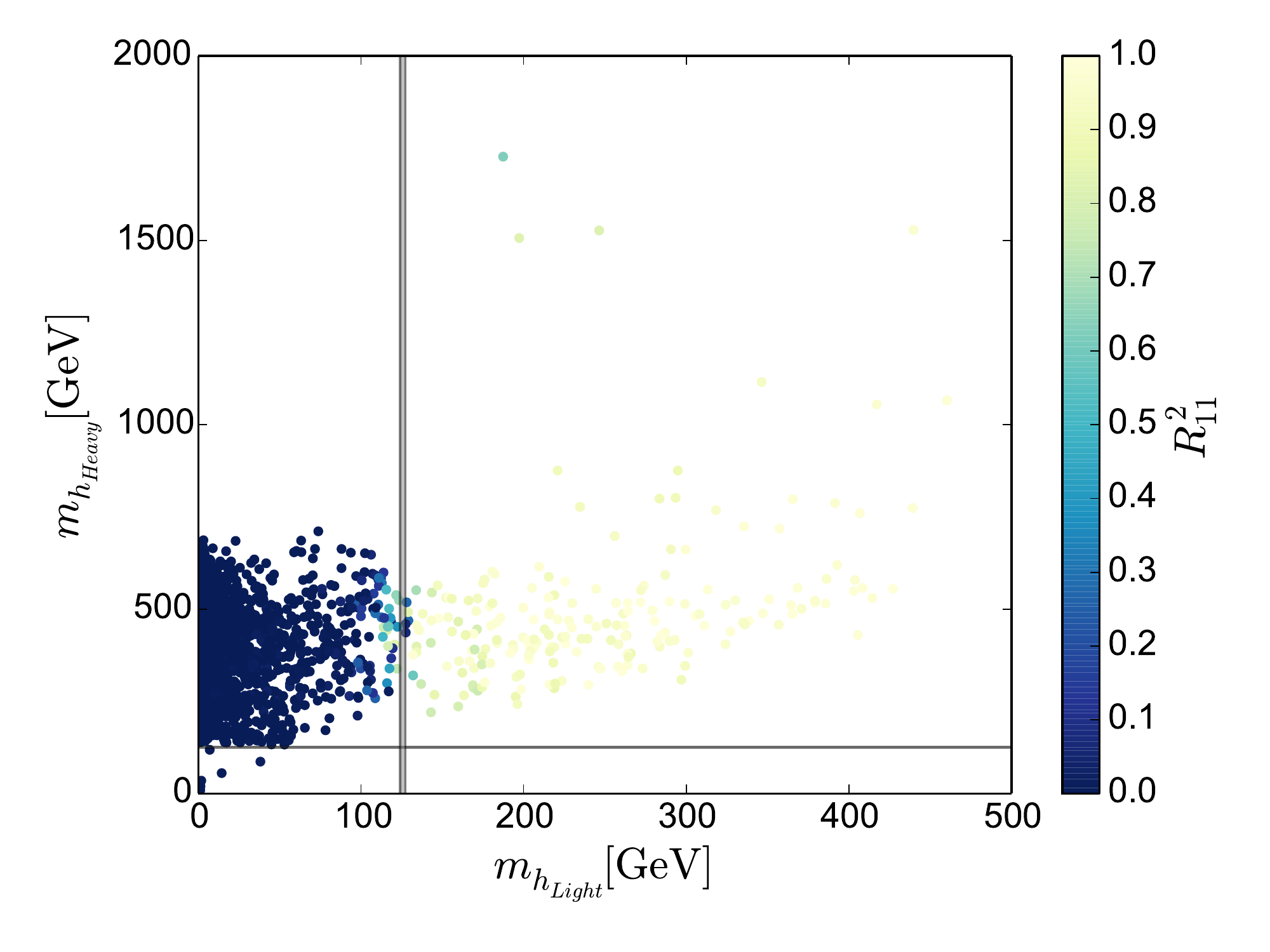}
  \caption{The $m_{h_{Light}} - m_{h_{Heavy}}$ plane with corresponding values of the squared Higgs mixing matrix elements $R_{11}^2$. All points shown pass the theoretical and experimental constraints of Section \ref{sec:scan}. The grey band highlights the SM Higgs mass range. Only points with $\beta_\lambda(M_{\rm Pl}), \beta_\delta(M_{\rm Pl}), \beta_{d_2}(M_{\rm Pl}) \leq 0.05$ are shown.}
\label{fig:complex_lighthiggs_heavyhiggs_Ri1}
\end{figure}

We note that Figure \ref{fig:complex_lmplanck_betal} contains a significant number of scenarios where the lightest extra scalar is considerably lighter than the SM Higgs boson. Since the model only couples the new scalars to the SM Higgs doublet, these scenarios escape detection at the LHC if the mixing with the SM Higgs is very small. The relevant quantity is $R_{11}^2$ (see Eq.~\ref{eq:complex_HBHS_signal_strength} and Appendix~\ref{Appendix:mixing}), which we show in Fig.~\ref{fig:complex_lighthiggs_heavyhiggs_Ri1}, demonstrating that the lightest scalar is indeed very decoupled in these scenarios.

We are in principle also interested in the high scale constraints $d_2 = \beta_{d_2} =0$ and/or $\delta=\beta_\delta=0$. However, we note that setting $\delta$ to zero at $M_{\rm Pl}$ decouples the extra scalars from the SM altogether, and since $\beta_\delta=0$ for this choice, $\delta$ remains zero at all scales and the new scalars are unobservable. However, as argued above, it is not unreasonable to consider $\delta$ {\em small} at the Planck scale, which is anyway phenomenologically necessary to keep the observed Higgs ``SM-like''. For non-zero values of $\delta$, it is also not possible to set $d_2$ exactly to zero at $M_{\rm Pl}$ since it is immediately driven negative by RG running and the vacuum destabilises according to \ref{eq:complex_vacuumstability2}. Again, we are forced to only consider $d_2$ {\em small} at the Planck scale and posit some new physics that causes this small deviation.

\begin{figure}[tbh]
  \centering
  \subfloat[]{\includegraphics[width=0.5\textwidth]{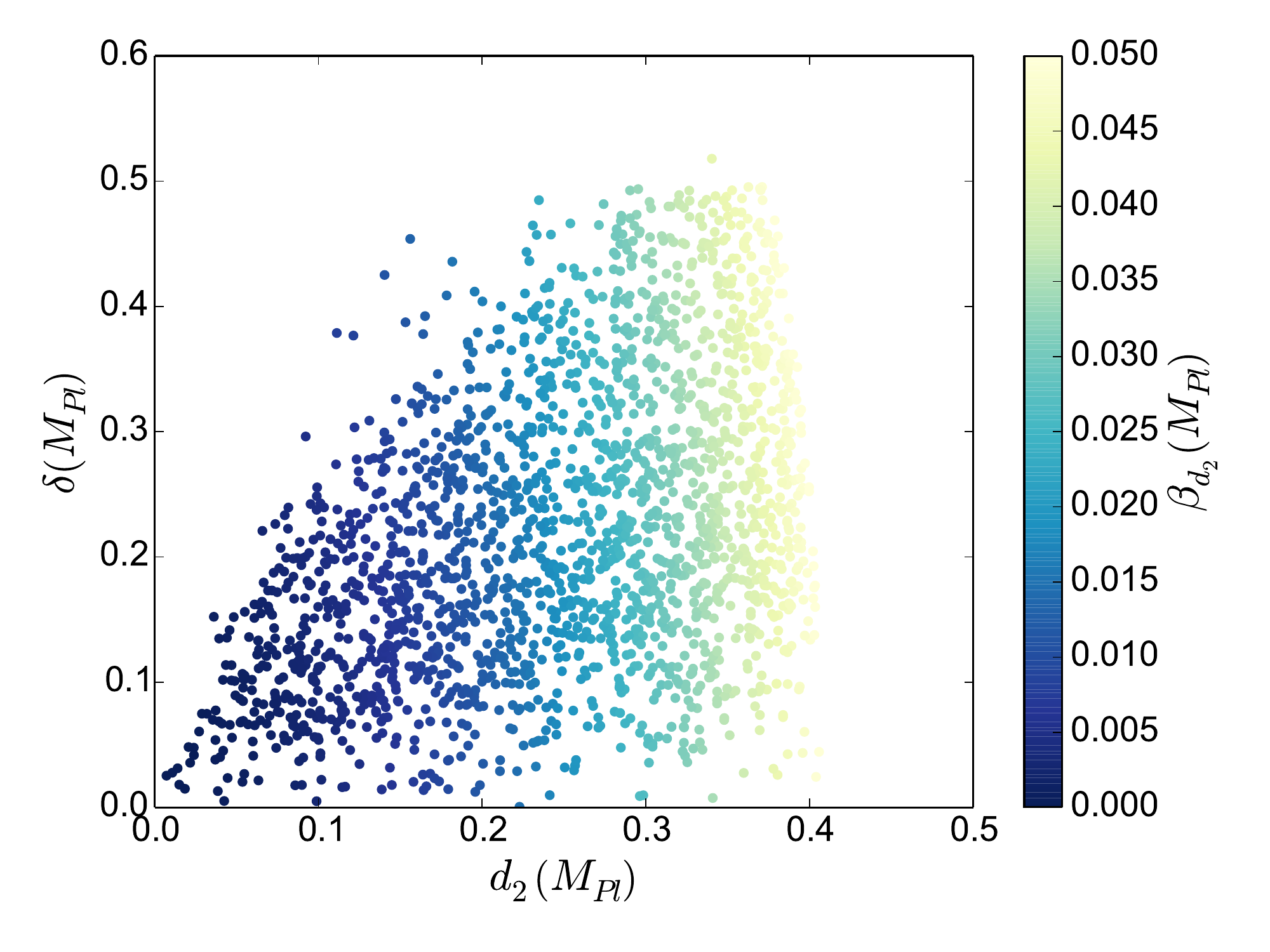}\label{fig:complex_d2_delta_betad2}}
  \hfill
  \subfloat[]{\includegraphics[width=0.5\textwidth]{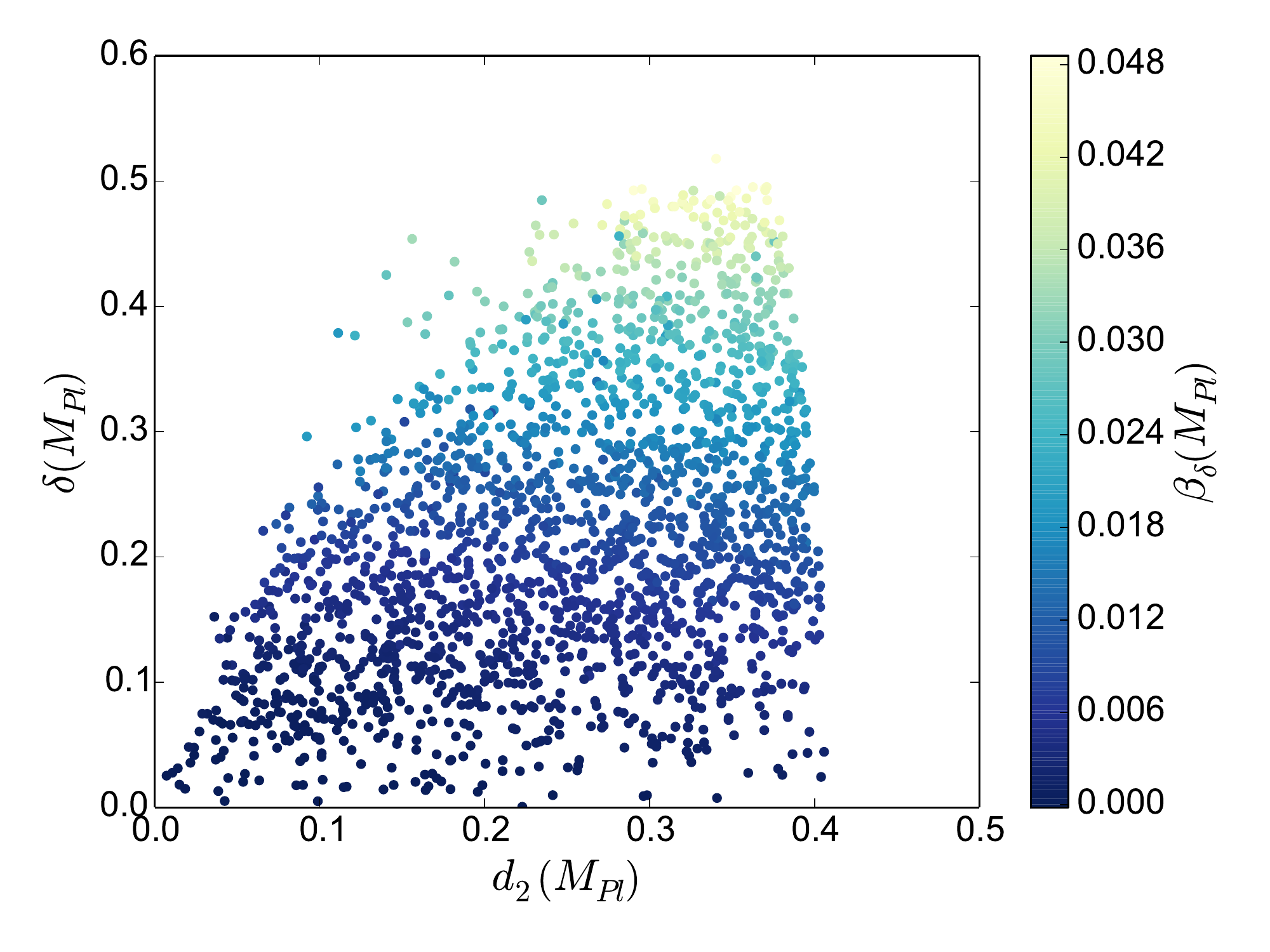}\label{fig:complex_d2_delta_betadelta}}
\caption{High scale Higgs quartic couplings $d_2$ and $\delta$ with their $\beta$-functions. Parameter points pass the theoretical and experimental constraints of Section~\ref{sec:scan}.}
\label{fig:complex_d2_delta_beta}
\end{figure}

In Figure \ref{fig:complex_d2_delta_beta} we show the space of allowed $d_2$ and $\delta$ and their high scale $\beta$ functions. As before, for clarity we exclude parameter points with $\beta$ functions larger than $0.05$ at $M_{\rm Pl}$. As for $\beta_\lambda$, the  vast majority of the valid points have $\beta_\delta$ lower than this cutoff, but we note that $\beta_{d_2}$ was able to be somewhat higher than $0.05$.

As one might expect, small values of the quartic couplings correspond with small values of their respective $\beta$ functions. It is also interesting to note that there exist valid scenarios that simultaneously have small values of both $\lambda, \delta, d_2$ and $\beta_{\lambda, \delta, d_2}$ at the Planck scale. These results make it clear that its possible to have a phenomenologically valid mass spectrum in the broken phase that is compatible with both the theoretical and experimental constraints whilst also allowing for  the boundary conditions $\lambda=\beta_\lambda=0$ to be at least {\em approximately} met.

\begin{figure}[!tbp]
  \centering
  \subfloat[]{\includegraphics[width=0.48\textwidth]{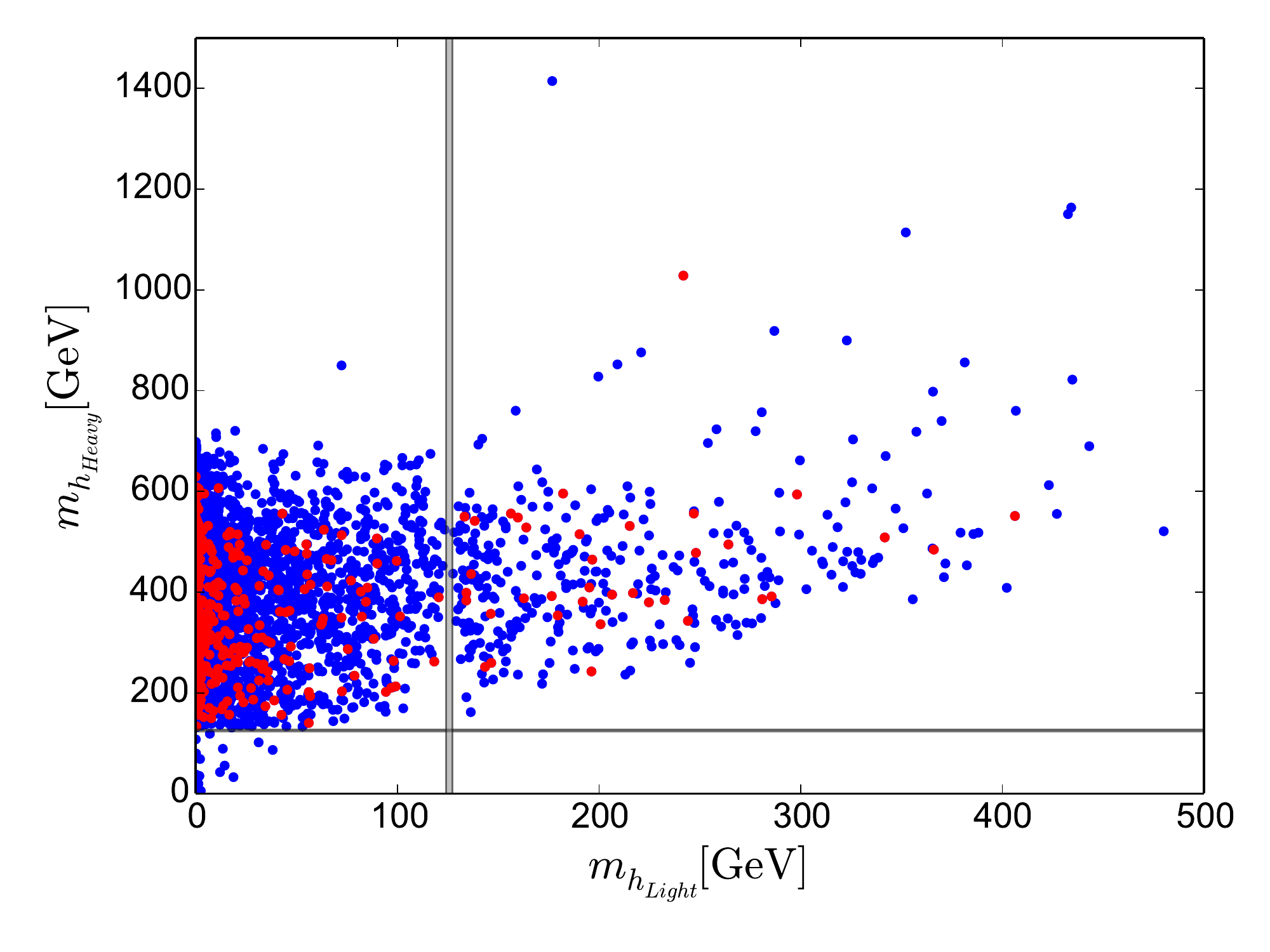}\label{fig:complex_lighthiggs_heavyhiggs_SMhiggs_comparison}}
  \hfill
  \subfloat[]{\includegraphics[width=0.48\textwidth]{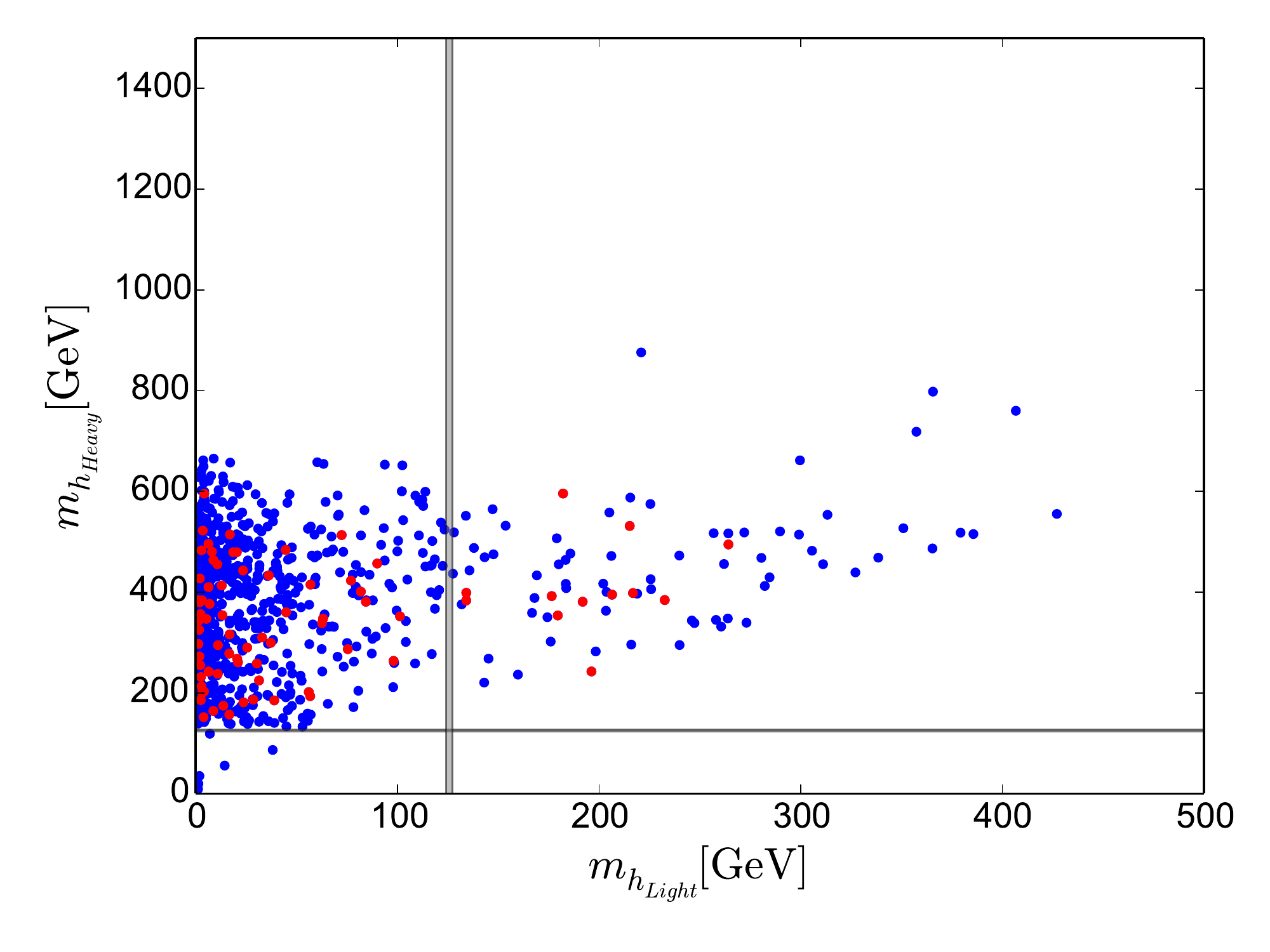}
 \label{fig:complex_lighthiggs_heavyhiggs_exp_comparison}}
\caption{Compatible values of $m_{h_{Light}}$ and $m_{h_{Heavy}}$ in the broken phase for different high scale $\beta_{\lambda}$ constraints. \textbf{(a)} includes points that are stable and perturbative up to $M_{Pl}$ and include an SM Higgs candidate, whilst \textbf{(b)} also enforces all experimental constraints. All points obey $\lambda<0.067$ at $M_{\rm Pl}$. Blue points obey $\beta_\lambda < 0.05$ at $M_{\rm Pl}$ while red points obey the more restrictive condition $\beta_\lambda < 0.00005$. The grey bands highlight the SM Higgs mass range.}
\label{fig:complex_lighthiggs_heavyhiggs_comparison}
\end{figure}

In Figure \ref{fig:complex_lighthiggs_heavyhiggs_comparison} we demonstrate the different scalar mass hierarchies in the $m_{h_{Light}} - m_{h_{Heavy}}$ plane, for small values of $\lambda$ and $\beta_\lambda$ at the high scale. The grey bands for the $SM$ Higgs mass cut the space into regions that have either two additional Higgs masses that are less than $m_{h_{SM}}$ (bottom-left quadrant), one less than $m_{h_{SM}}$ and one heavier (top-left), or two heavier additional scalars (top-right). The effect of the experimental constraints described in Section~\ref{sec:scan} can be seen by comparing Figures \ref{fig:complex_lighthiggs_heavyhiggs_comparison}(a) and  \ref{fig:complex_lighthiggs_heavyhiggs_comparison}(b). In Figure \ref{fig:complex_lighthiggs_heavyhiggs_comparison}(a) we apply only theoretical constraints (such as vacuum stability), while in \ref{fig:complex_lighthiggs_heavyhiggs_comparison}(b) we also apply the experimental bounds. Irrespectively, the majority of valid points fall into the top-left or top-right quadrants. The blue points respect only the (unrestrictive) bound $\beta_{\lambda} <  0.05$, while red points have $\beta_\lambda < 0.00005$ and are therefore consistent with zero. The smaller cutoff excludes all of the points with two scalars lighter than $m_{h_{SM}}$ and most of the points where the SM Higgs is the lightest of the three. Indeed, the lighter additional scalar mass never exceeds about $260\,$GeV when the more restrictive cutoff is used. The heavier scalar never falls below $\sim 140\,$GeV and never exceeds $\sim 800 \,$GeV.

\begin{figure}[!tbp]
  \centering
  \subfloat[]{\includegraphics[width=0.5\textwidth]{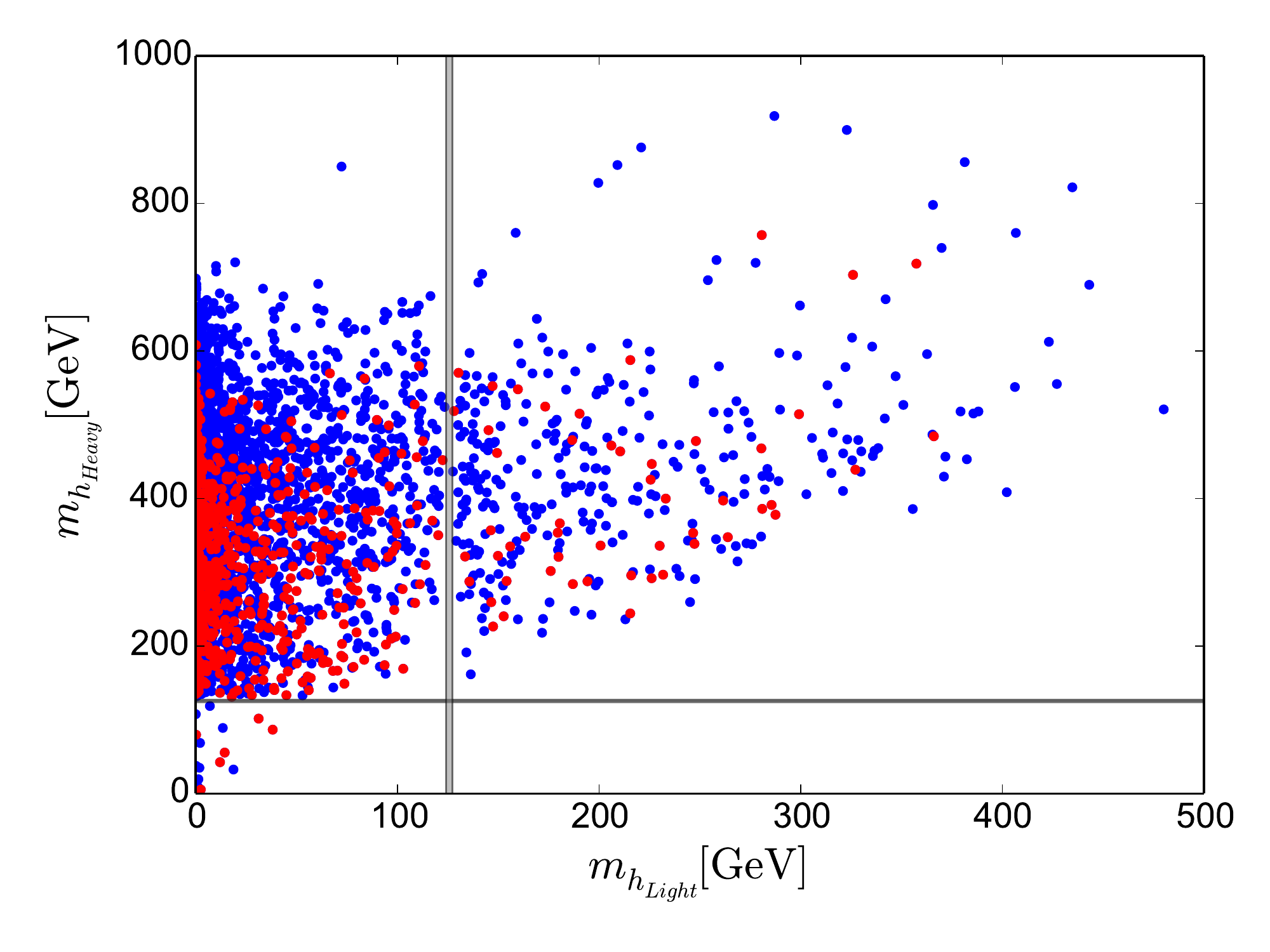}\label{fig:complex_lighthiggs_heavyhiggs_SMhiggs_comparison_10times}}
  \hfill
  \subfloat[]{\includegraphics[width=0.5\textwidth]{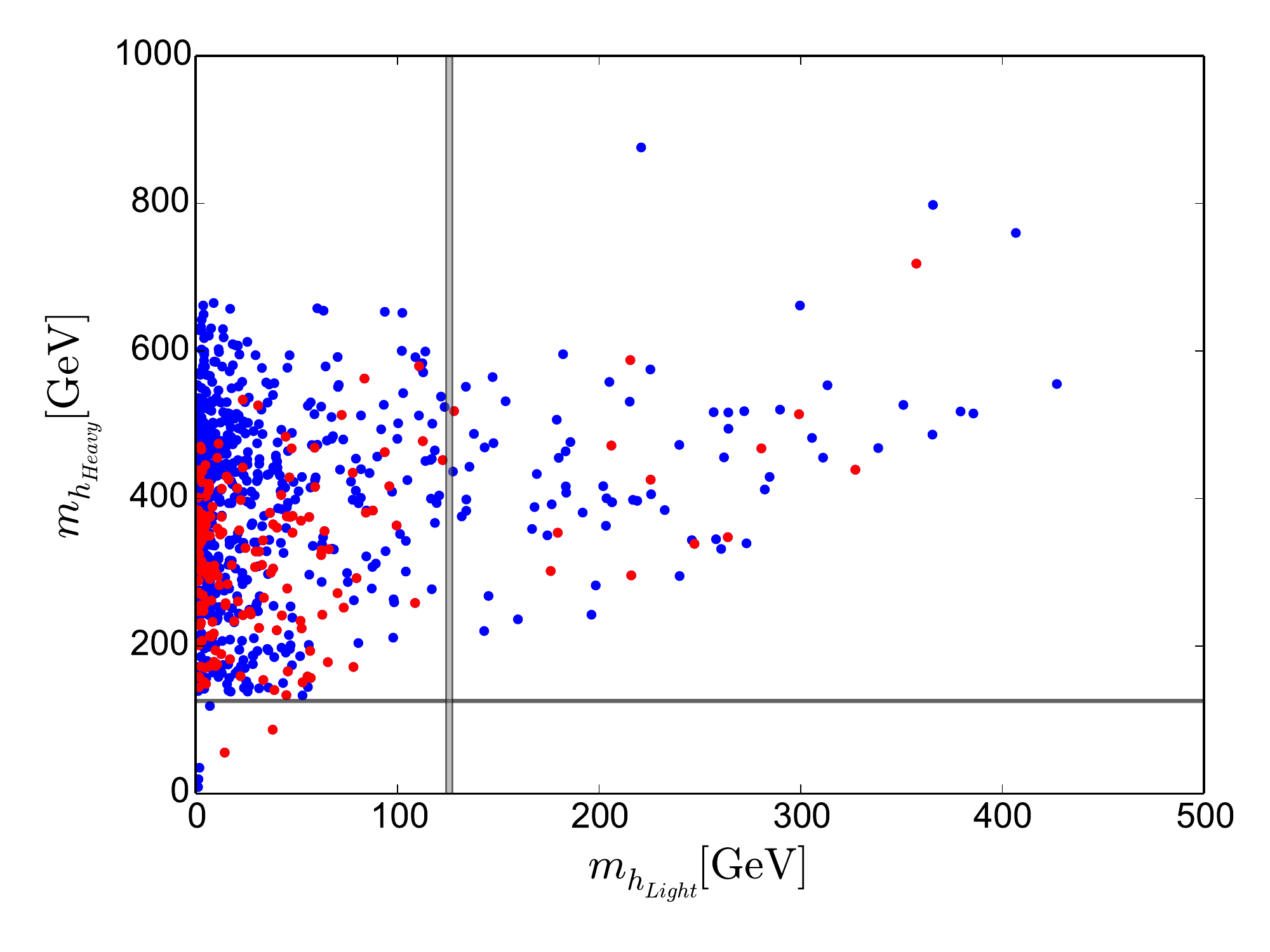}\label{fig:complex_lighthiggs_heavyhiggs_exp_comparison_10times}}
\caption{Compatible values of $m_{h_{Light}}$ and $m_{h_{Heavy}}$ in the broken phase with restrictions on $\beta_{\lambda}$, $\beta_\delta$ and $\beta_{d_2}$. \textbf{(a)} includes points that are stable and perturbative up to $M_{Pl}$ and include an SM Higgs candidate, whilst \textbf{(b)} also enforces all experimental constraints. All points obey $\lambda<0.067$ at $M_{\rm Pl}$. Blue points obey $\beta_{\lambda,\delta,d_2} < 0.05$ at $M_{\rm Pl}$ while red points obey $\beta_\lambda < 0.0005$, $\beta_\delta<0.0025$ and $\beta_{d2}<0.01$. The grey bands highlight the SM Higgs mass range.}
\label{fig:complex_lighthiggs_heavyhiggs_comparison_10times}
\end{figure}

Using the difference between one- and two-loop running to estimate what constitutes ``small'' for the $\beta$-functions of $\delta$ and $d_2$, we find the constraints,
\begin{eqnarray}
\beta_{\delta} \left( M_{Pl} \right) &\lesssim & 0.00025, \nonumber \\
\beta_{d_2} \left( M_{Pl} \right) &\lesssim & 0.001.
\label{eq:complex_DM_truncation_error}
\end{eqnarray}
No broken-phase parameter points survive if we include all three of the tightest $\beta$-function constraints simultaneously in addition to the experimental constraints, indicating that strictly enforcing the conditions \ref{eq:complex_bc1} and \ref{eq:complex_bc2} is incompatible with experiment. However, if the boundary conditions are imposed by some new UV theory or principle, it may be that new physics exists at or around $M_{\rm Pl}$ that distorts the running of the quartics as we approach. Without knowing the form of this UV completion, we don't know the size of these threshold corrections, so don't know how much deviation from zero we should allow in our boundary conditions. With this in mind we may regard these constraints as too conservative. To investigate their loosening, we somewhat arbitrarily relax our boundary condition cut-offs to ten times those of  \ref{eq:betal_cutoff} and \ref{eq:complex_DM_truncation_error}. We now find the points survive and plot these in figure \ref{fig:complex_lighthiggs_heavyhiggs_comparison_10times}. Notice that this also loosens of the constraint on $\beta_{\lambda} \left( M_{Pl} \right)$ used in figure \ref{fig:complex_lighthiggs_heavyhiggs_comparison} and now a small number of points survive that have the SM Higgs as the heaviest of the three scalars.

\section{The Dark Matter Phase}
\label{sec:results_DM}

\begin{figure}[h]
  \centering
  \subfloat[]{\includegraphics[width=0.5\textwidth]{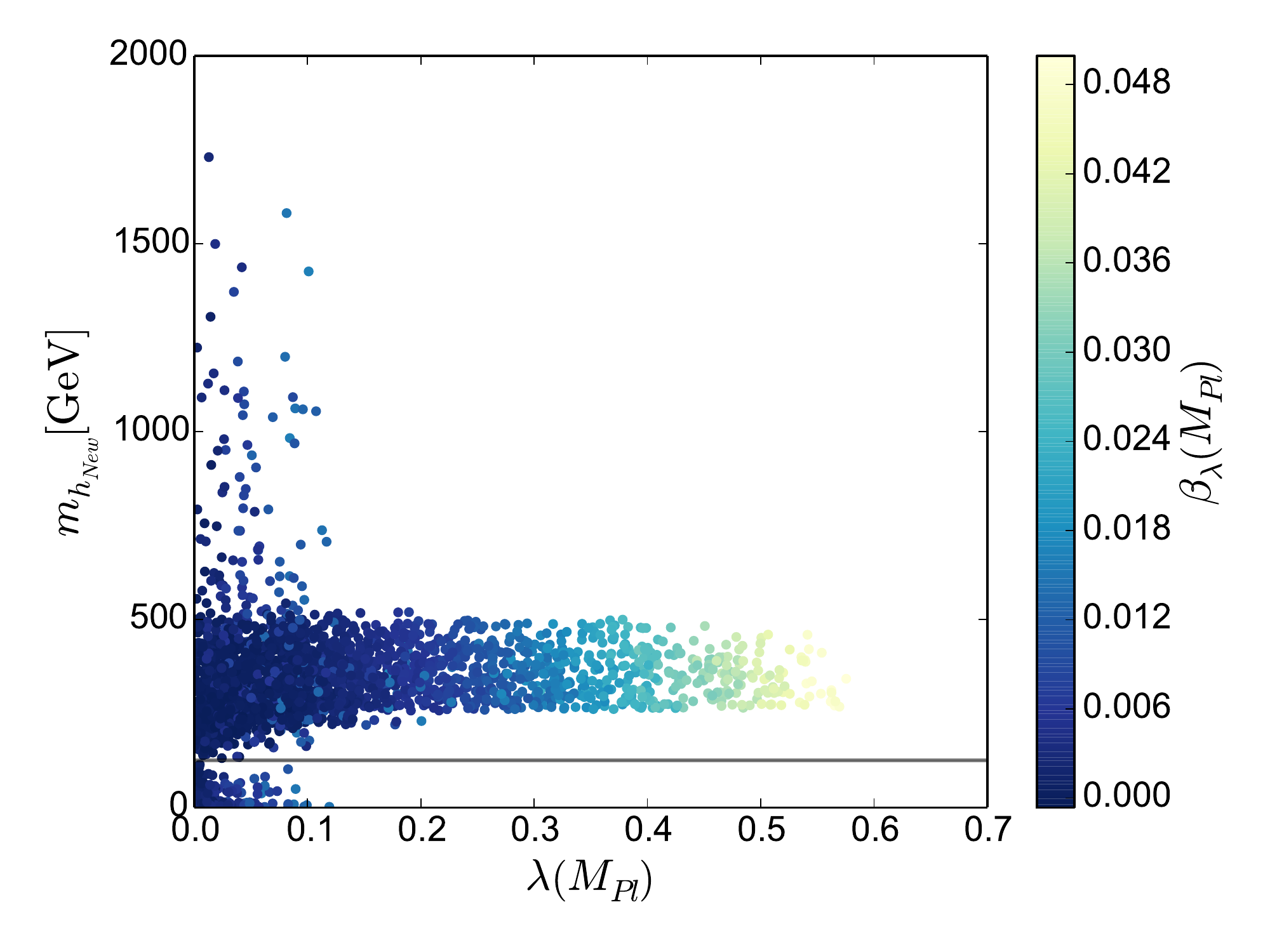}\label{fig:complex_DM_lmplanck_additionalhiggs_betal}}
  \hfill
  \subfloat[]{\includegraphics[width=0.5\textwidth]{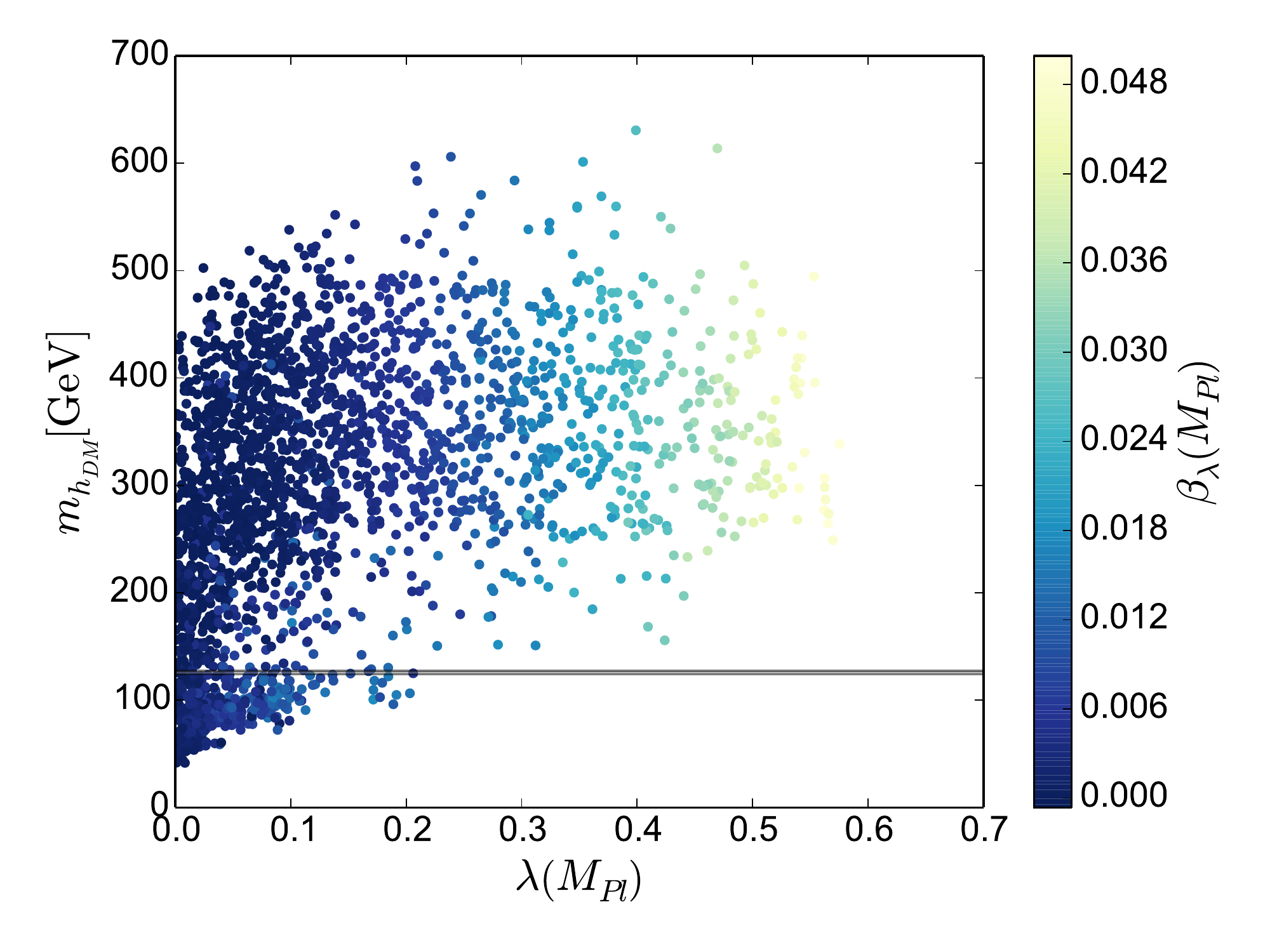}\label{fig:complex_DM_lmplanck_DMhiggs_betal}}
\caption{High scale Higgs quartic coupling $\lambda$ vs.\ the additional Higgs mass $m_{h_{New}}$ or the DM scalar mass $m_{h_{DM}}$, with values of $\beta_\lambda$. Parameter points pass the theoretical and experimental constraints of Section~\ref{sec:scan}, including dark matter relic density and direct detection constraints. The grey band shows the SM Higgs mass range.}
\label{fig:complex_DM_lmplanck_betal}
\end{figure}

\begin{figure}[!tbp]
  \centering
  \subfloat[]{\includegraphics[width=0.5\textwidth]{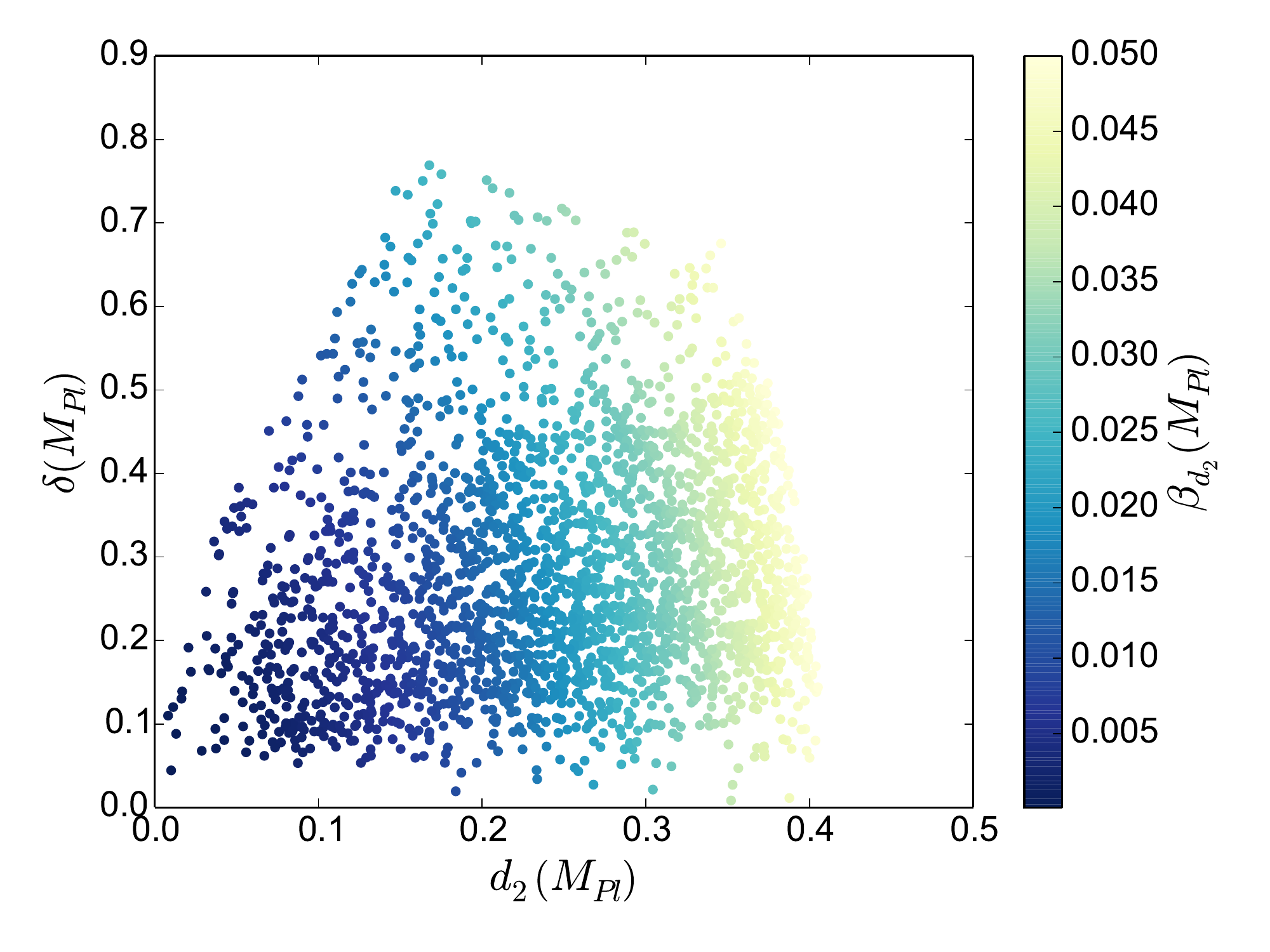}\label{fig:complex_DM_d2_delta_betad2}}
  \hfill
  \subfloat[]{\includegraphics[width=0.5\textwidth]{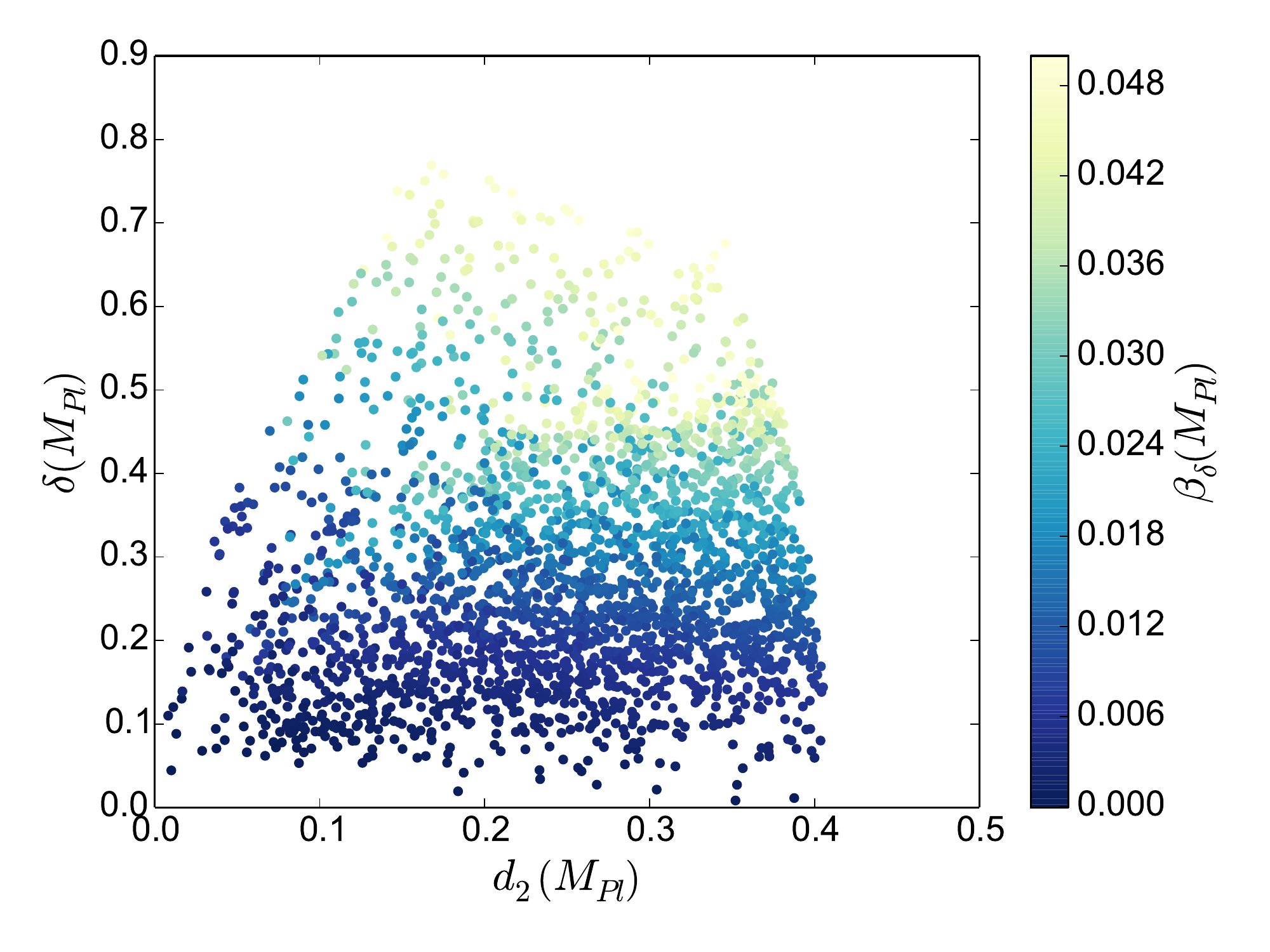}\label{fig:complex_DM_d2_delta_betadelta}}
\caption{High scale Higgs quartic couplings $d_2$ and $\delta$ with their corresponding $\beta$-functions, in the DM phase. Parameter points pass the theoretical and experimental constraints of Section~\ref{sec:scan}, including dark matter relic density and direct detection constraints.}
\label{fig:complex_DM_d2_delta}
\end{figure}

In the dark matter phase only two of the three scalars are allowed to mix, with the third becoming a dark matter candidate. We call the non-SM-like Higgs as $h_{\rm New}$ whilst the DM scalar is $h_{DM}$. Figure \ref{fig:complex_DM_lmplanck_betal} shows  high scale $\lambda$ vs.\ either $m_{h_{New}}$ or $m_{h_{DM}}$, including theoretical and experimental constraints, as well as each point's corresponding value of $\beta_{\lambda} \left( M_{\rm Pl} \right)$. Figure \ref{fig:complex_DM_d2_delta} shows the valid ranges of $d_2$ and $\delta$ as well as their respective $\beta$-functions at $M_{\rm Pl}$. These figures are analogous to Figures \ref{fig:complex_lmplanck_betal} and \ref{fig:complex_d2_delta_beta}, and again for clarity we are restricting the $\beta$-functions at $M_{\rm Pl}$ to be smaller than $0.05$.

In contrast to the broken phase, most of the valid points have a additional Higgs $m_{h_{\rm New}}$ greater than the SM Higgs mass, illustrated by the grey horizontal band, with the majority of those points falling into a range between around $m_{h_{SM}}$ and approximately $500\,$GeV. It's interesting to note that the points that do result in $m_{h_{New}} < m_{h_{SM}}$ have smaller values of $\lambda \left( M_{\rm Pl} \right) \lesssim 0.1$. As was the case in the broken phase, smaller values of the quartic couplings correspond to smaller $\beta$ functions. The dark matter candidate mass $m_{h_{DM}}$ has a lower limit of about $40\,$GeV, as can be seen in Figure \ref{fig:complex_DM_lmplanck_betal}(b), which is in keeping with the results of \citep{Costa:2014qga}. Here, however, we point out that points at this low end of the mass range also have small values of both $\beta_{\lambda} \left( M_{Pl} \right)$ and $\beta_{\delta} \left( M_{Pl} \right)$.

\begin{figure}[!tbp]
  \centering
  \subfloat[]{\includegraphics[width=0.5\textwidth]{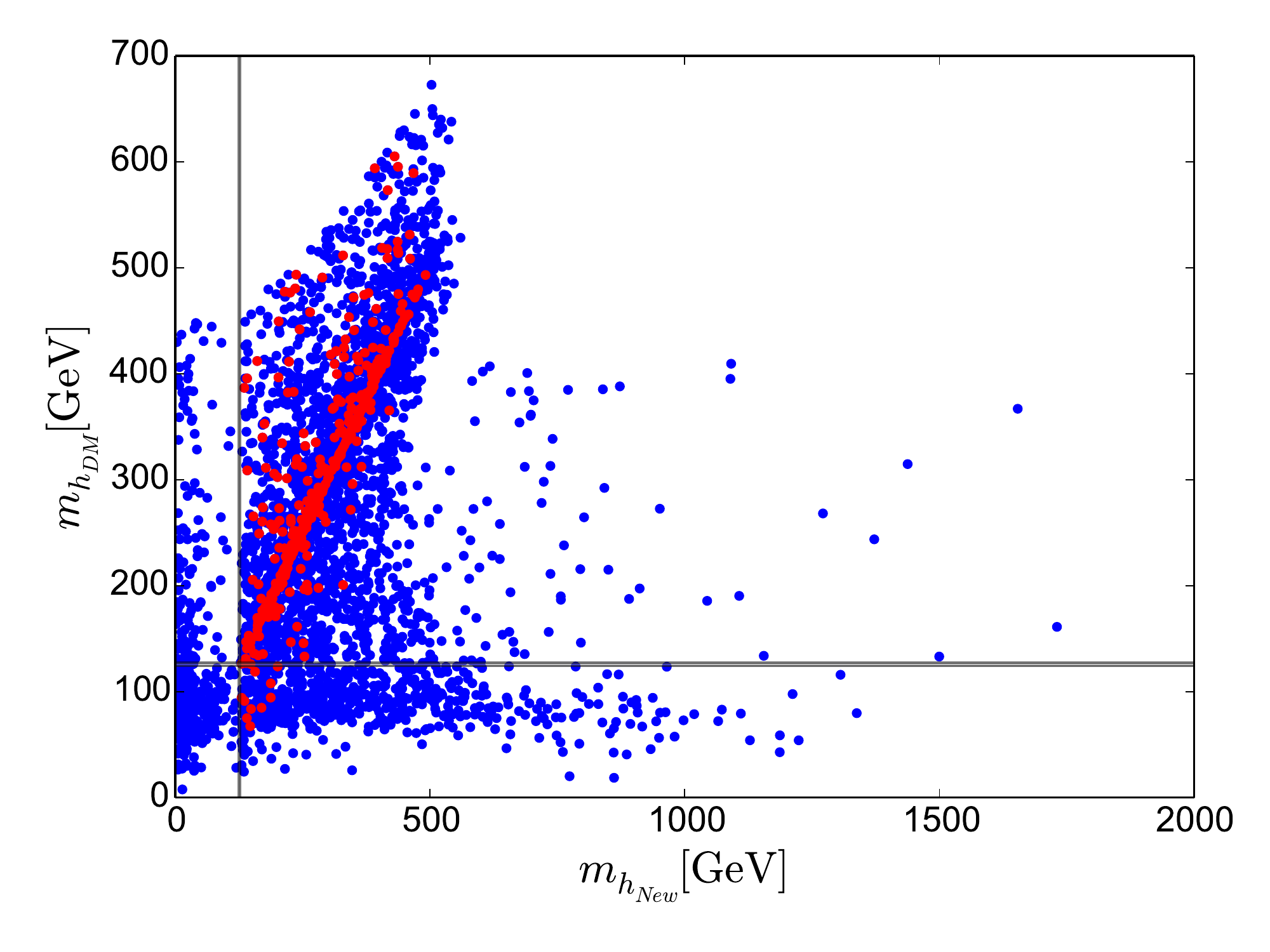}\label{fig:complex_DM_additionalhiggs_DMhiggs_SMhiggs_comparison}}
  \hfill
  \subfloat[]{\includegraphics[width=0.5\textwidth]{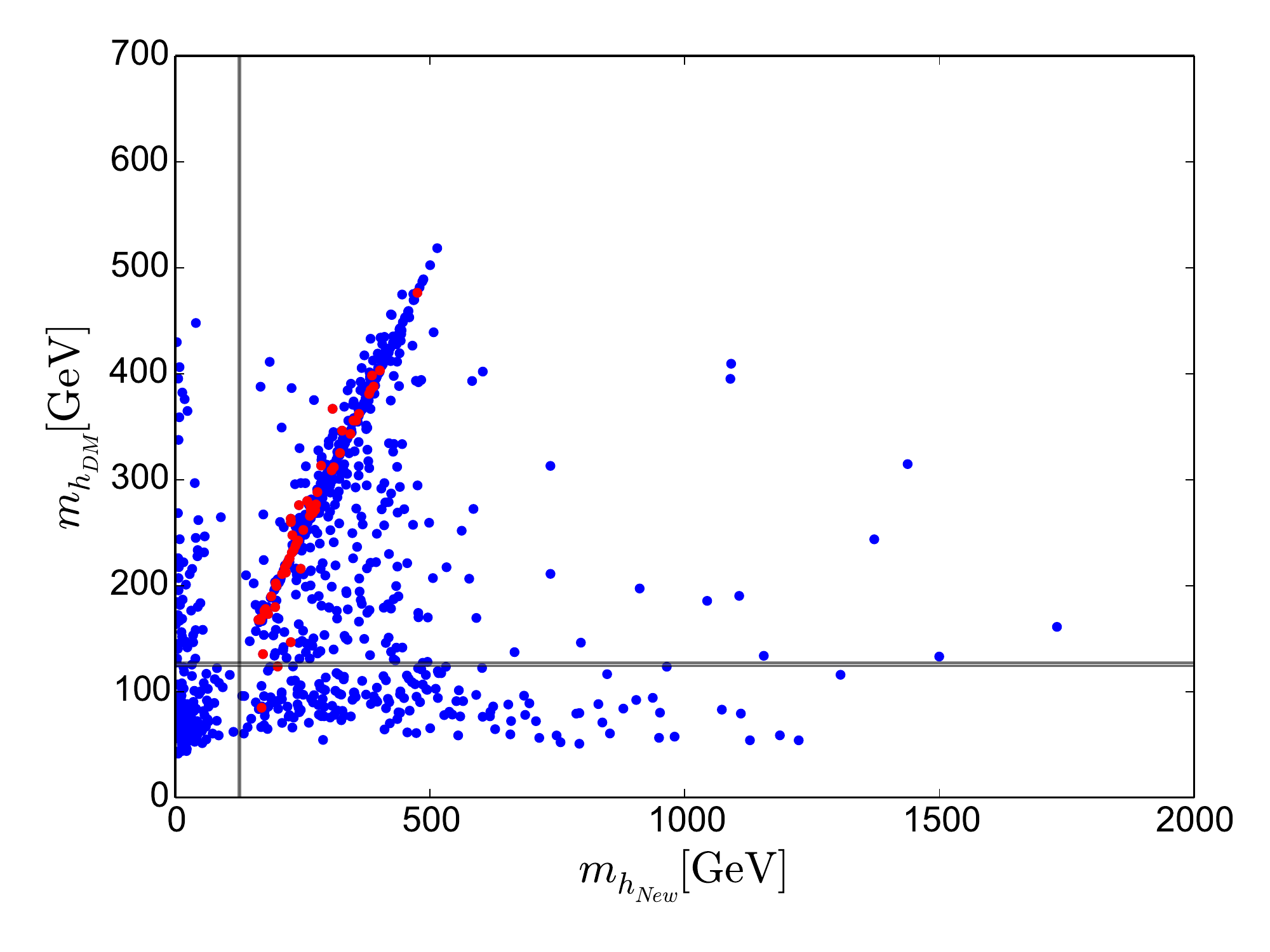}\label{fig:complex_DM_additionalhiggs_DMhiggs_exp_comparison}}
\caption{Compatible values of $m_{h_{New}}$ and $m_{h_{DM}}$ in the DM phase for different high scale $\beta_{\lambda}$ constraints. \textbf{(a)} includes points that are stable and perturbative up to $M_{Pl}$ and include an SM Higgs candidate, whilst \textbf{(b)} also enforces all experimental constraints. All points obey $\lambda<0.067$ at $M_{\rm Pl}$. Blue points obey $\beta_\lambda < 0.05$ at $M_{\rm Pl}$ while red points obey the more restrictive condition $\beta_\lambda < 0.00005$. The grey bands highlight the SM Higgs mass range.}
\label{fig:complex_DM_additionalhiggs_DMhiggs_comparison}
\end{figure}

Figure \ref{fig:complex_DM_additionalhiggs_DMhiggs_comparison} examines the extra scalar masses when we restrict $\lambda$ and $\beta_\lambda$ to be consistent with zero.  Again, for comparison, we show points with a very unrestrictive $\beta_\lambda<0.05$ is blue before demonstrating the effect of the constraint $\beta_\lambda<0.00005$ in red. No points with $m_{h_{New}} < m_{h_{SM}}$ survive the stronger constraint on $\beta_{\lambda}$, and the majority of the points that do survive have almost degenerate masses of $m_{h_{New}}$ and $m_{h_{DM}}$. The tree level masses of $m_{h_{New}}$ ($m_{h_{DM}}$) have a linear dependence on $a_1$ ($b_{-}$) which appears to dominate when both of the additional scalars are heavier than the SM Higgs (the mass matrix is described in Appendix~\ref{Appendix:mixing}). This degeneracy is visible in the parameter space where only theoretical constraints are applied but is much more pronounced when the experimental constraints are also in place, where much of the parameter space is ruled out primarily via the WMAP and Planck relic density constraint. There is a lower limit $m_{h_{New}} \gtrsim 130\,$GeV if we include only the theoretical constraints, which rises to $\gtrsim 160\,$GeV if we include experimental constraints. This lower limit in $m_{h_{New}}$ is similar to the lower limit on $m_{h_{Heavy}}$ in the broken phase that we discussed in Section \ref{sec:results_broken}.

\begin{figure}[tbh]
  \centering
  \subfloat[]{\includegraphics[width=0.5\textwidth]{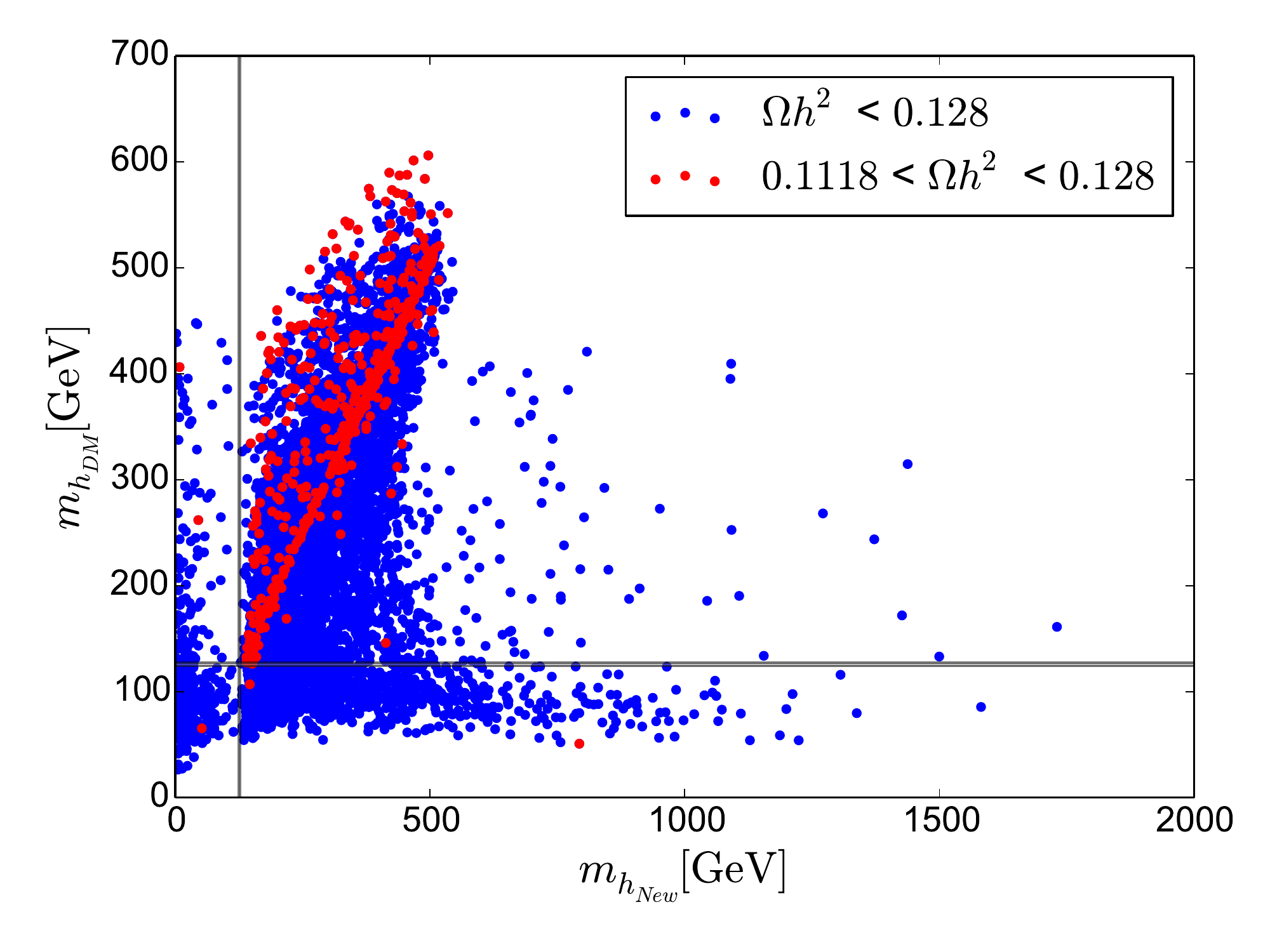}}
	\hfill
	\subfloat[]{\includegraphics[width=0.5\textwidth]{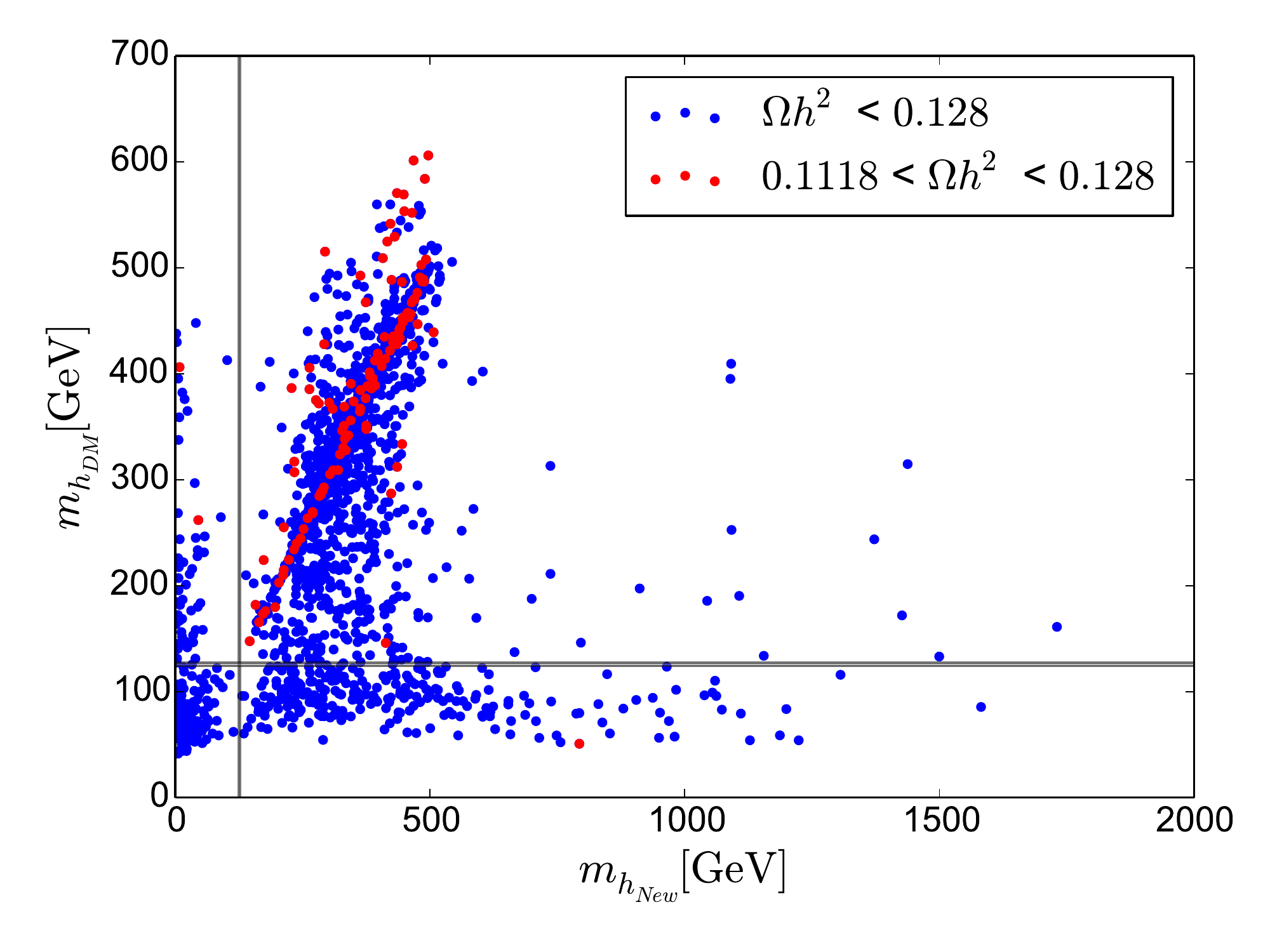}}
  \caption{Comparison of dark matter relic density $\Omega h^2$ in the $m_{h_{New}} - m_{h_{DM}}$ plane. Blue points result in $\Omega h^2 < 0.128$ whilst the red points satisfy the stronger constraint $0.1118 < \Omega h^2 < 0.128$. \textbf{(a)} includes points that are stable and perturbative up to $M_{Pl}$ and include an SM Higgs candidate, whilst \textbf{(b)} also enforces all experimental constraints. The grey band highlights the SM Higgs mass range.}
\label{fig:complex_DM_additionalhiggs_DMhiggs_omega_comparison}
\end{figure}

Looking at figure \ref{fig:complex_DM_additionalhiggs_DMhiggs_comparison} might suggest that small values of the $\beta$ functions at the Planck scale correlates with a small mass difference $\Delta m = |m_{h_{New}} - m_{h_{DM}}|$. However, while $80\%$ of the points that pass through the constraint $\lambda<0.067, \beta_\lambda < 0.00005$ (red points) result in $\Delta m < 40$ GeV, so do $67\%$ of the (blue) points that don't. This tendency towards degeneracy is a feature of all of the points that satisfy the theoretical constraints outlined in section \ref{sec:scan}. These points exhibit small values of the soft $U(1)$ breaking parameters $a_1$ and $b_1$, forcing a small $\Delta m$ \citep{Coimbra:2013qq}. It is interesting to note that many points in the degenerate mass region can completely account for the dark matter relic density, as shown in figure \ref{fig:complex_DM_additionalhiggs_DMhiggs_omega_comparison}. The degeneracy opens up co-annihilation channels involving both $m_{h_{DM}}$ and $m_{h_{New}}$ that enter the relic density calculation \citep{Baker:2015qna,Ghorbani:2014gka}. These new channels help bring down the relic density to within the $3 \sigma$ range.

\begin{figure}[!tbp]
  \centering
  \subfloat[]{\includegraphics[width=0.5\textwidth]{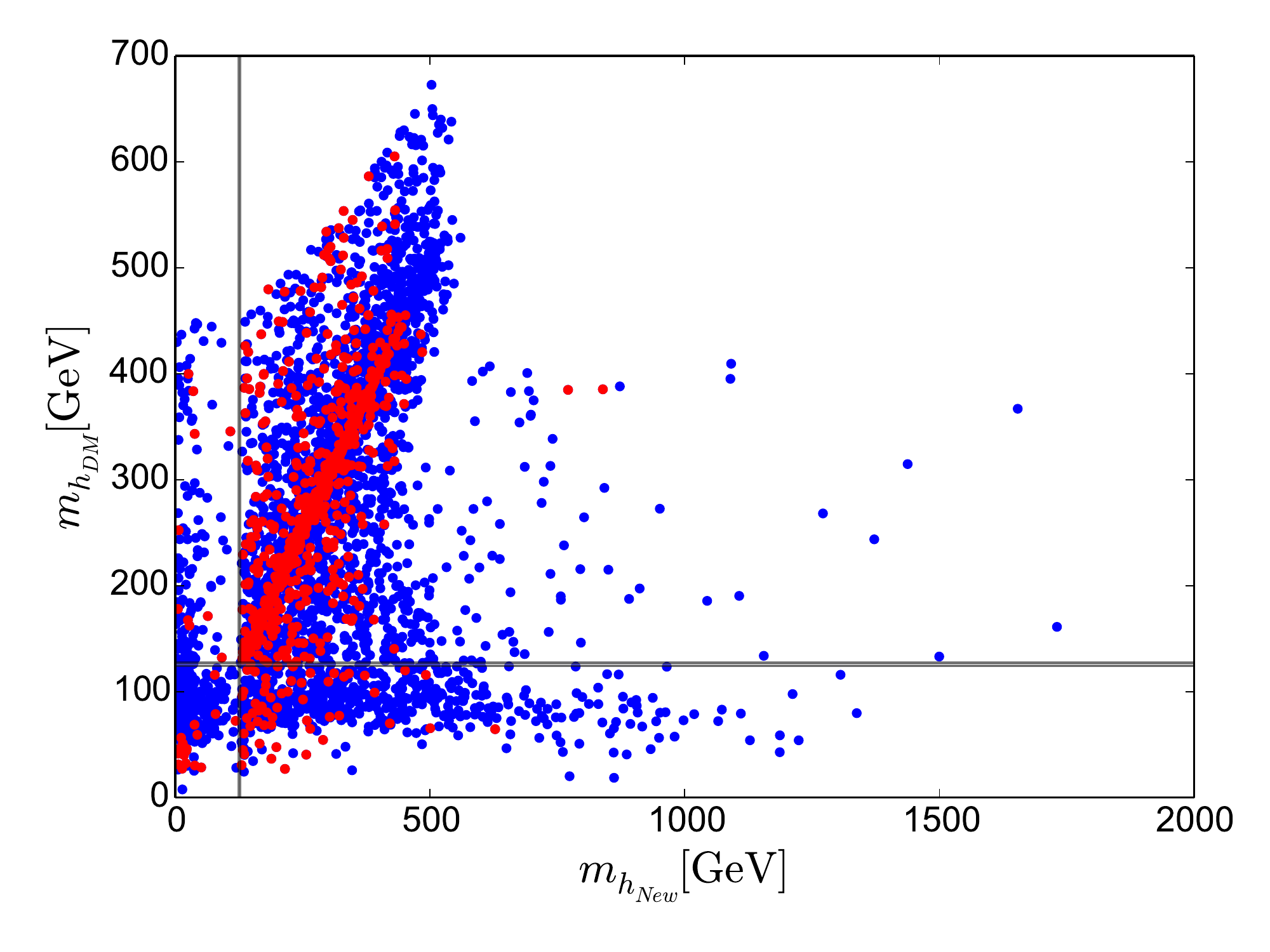}\label{fig:complex_DM_additionalhiggs_DMhiggs_SMhiggs_comparison_10times}}
  \hfill
  \subfloat[]{\includegraphics[width=0.5\textwidth]{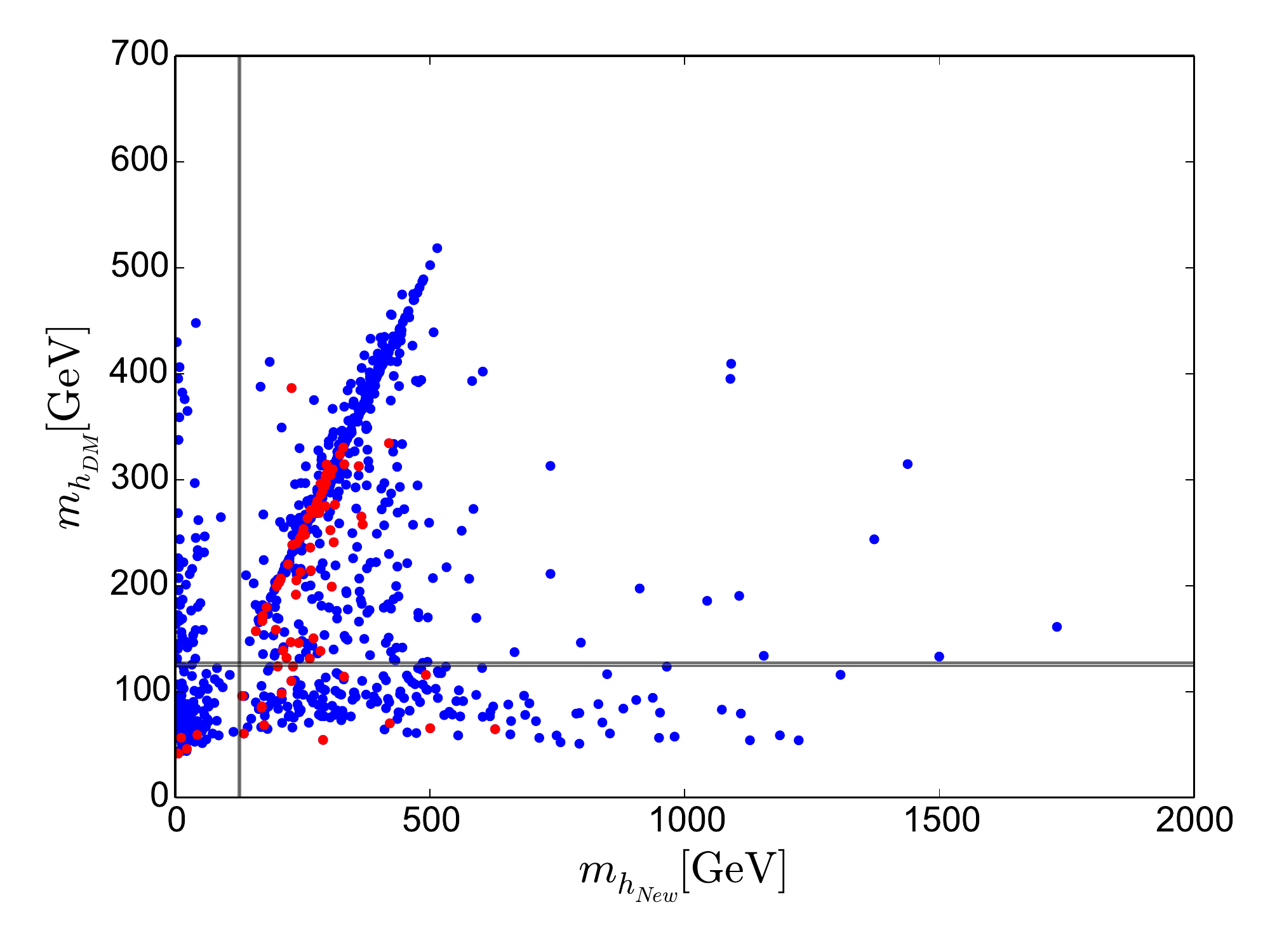}\label{fig:complex_DM_additionalhiggs_DMhiggs_exp_comparison_10times}}
\caption{Compatible values of $m_{h_{New}}$ and $m_{h_{DM}}$ in the DM phase with restrictions on $\beta_{\lambda}$, $\beta_\delta$ and $\beta_{d_2}$. \textbf{(a)} includes points that are stable and perturbative up to $M_{Pl}$ and include an SM Higgs candidate, whilst \textbf{(b)} also enforces all experimental constraints. All points obey $\lambda<0.067$ at $M_{\rm Pl}$. Blue points obey $\beta_{\lambda,\delta,d_2} < 0.05$ at $M_{\rm Pl}$ while red points obey $\beta_\lambda < 0.0005$, $\beta_\delta<0.0025$ and $\beta_{d2}<0.01$. The grey bands highlight the SM Higgs mass range.}
\label{fig:complex_DM_additionalhiggs_DMhiggs_comparison_10times}
\end{figure}

As in the broken phase, no DM phase points survive when \ref{eq:betal_cutoff} and \ref{eq:complex_DM_truncation_error} are applied simultaneously with the experimental constraints.
However, we see scenarios survive if we relax the constraints by a factor of 10. These scenarios are shown in Figure \ref{fig:complex_DM_additionalhiggs_DMhiggs_comparison_10times}.

% \begin{figure}[!tbp]
%   \centering
%   \subfloat[]{\includegraphics[width=0.5\textwidth]{TwoSSMbpmaDM_additionalhiggs_DMhiggs_a1_cuberoot.pdf}\label{fig:complex_DM_additionalhiggs_DMhiggs_a1_cuberoot}}
%   \hfill
%   \subfloat[]{\includegraphics[width=0.5\textwidth]{TwoSSMbpmaDM_additionalhiggs_DMhiggs_bminus_sqrt.pdf}\label{fig:complex_DM_additionalhiggs_DMhiggs_bminus_sqrt}}
% \caption{Points in the $m_{h_{New}} - m_{h_{DM}}$ plane and their corresponding values of \textbf{(a)} soft linear term $a_{1}^{\frac{1}{3}}$ and \textbf{(b)} bilinear $\sqrt{b_{-}}$, in the DM phase. Parameter points pass the theoretical constraints discussed in Section~\ref{sec:scan}.}
% \label{fig:complex_DM_additionalhiggs_DMhiggs_a1_bminus}
% \end{figure}

\section{Conclusions}
\label{sec:conclusions}
We have examined the Complex Singlet extension of the Standard Model with additional constraints on the model's quartic couplings at the Planck scale. These boundary conditions may arise due to high scale requirements of the potential in the Multiple Point Principle, or the evolution towards a UV fixed point, as in the Asymptotic Freedom or Asymptotic Safety scenarios. Here we have not focused on the precise mechanism by which these conditions may have arisen, but examine the boundary conditions themselves to see if they are compatible with theoretical constraints and experimental observations.

The model exhibits multiple phases, including a ``broken'' phase, in which both real and imaginary parts of the extra singlet gain a vev and the three scalars all mix; and the ``Dark Matter'' phase, in which only the real part of the extra singlet gains a vev, only two scalars mix and the remaining scalar provides a stable dark matter candidate.

We investigate a wide range of parameter space in both phases, rejecting parameter choices that do not provide a SM-like Higgs with mass $\approx 125$ GeV. We also apply theoretical constraints, such as vacuum stability and perturbativity up to the Planck scale $M_{Pl}$, and further experimental constraints such as Higgs production and decay rates, and where appropriate constraints on the Dark Matter relic density. We then examine the Planck scale values of the Higgs quartic couplings and their corresponding $\beta$ functions.

The addition of the complex singlet gives considerable flexibility for imposition of the boundary conditions $\lambda= \beta_\lambda=0$ at the Planck Scale. Indeed, we find regions of parameter space in both the broken and DM phases where this boundary condition is realised while maintaining compatibility with current theoretical and experimental constraints. Scenarios with {\em all} quartic couplings, including that of the additional scalar and the Higgs portal interaction, {\em exactly} zero are not possible since the Higgs portal never regenerates with RGE running once it is set to zero at $M_{\rm Pl}$. However, if some new physics theory at the high scale makes it simply {\em very small}, then compatibility with all low energy observations can be restored.

In the broken phase we found that the majority of valid scenarios have one additional Higgs that is lighter than the SM Higgs and one that is heavier. In contrast, in the DM phase most scenarios have additional scalars that are heavier than the SM-like Higgs and degenerate with one another. These scalars are all rather decoupled and difficult to detect, but could possibly be investigated at the high luminosity run of the LHC or at future colliders \citep{Chang:2017ynj}. Imposing $\lambda= \beta_\lambda=0$ requires the heaviest additional scalar in the broken phase to be lighter than about $600\,$GeV, with a lower upper limit of $\sim 500\,$GeV in the DM phase. These results are in keeping with previous work on vacuum stability in the complex singlet model \citep{Costa:2014qga}.

An interesting path for future work would be to investigate similar boundary conditions in theories with additional field content. For example, how is this picture changed in a two-Higgs-doublet model, or in theories containing vector-like fermions?

\section*{Acknowledgements}

The authors would like to thank Peter Athron for invaluable help with FlexibleSUSY; as well as Karl Nordstrom, Ant\'onio Morais and David Sutherland for useful discussions. DJM acknowledges partial support from the STFC grants ST/L000446/1 and ST/P000746/1.

\begin{appendices}
\section{Mass Matrices}
\label{Appendix:mixing}
The mass eigenstates $H = \left( h_1, h_2, h_3 \right)$ are related to the gauge eigenstates $\rho = \left( h, s_1, s_2 \right)$ by a 3x3 rotation matrix $R$,
\begin{equation}
\label{eq:complex_rotation}
H_i = R_{ij} \rho_j,
\end{equation}
with
\begin{equation}
\label{eq:complex_rotation_matrix}
R =
\begin{pmatrix}
c_1 c_2 & s_1 c_2 & s_2 \\
-( c_1 s_2 s_3 + s_1 c_3) & c_1 c_3 - s_1 s_2 s_3 & c_2 s_3 \\
- c_1 s_2 c_3 + s_1 s_3 & -( c_1 s_3 + s_1 s_2 c_3) & c_2 c_3
\end{pmatrix},
\end{equation}
where $s_i \equiv \sin \alpha_i$, $c_i \equiv \cos \alpha_i$ and $ \left| \alpha_i \right| \leq \frac{\pi}{2}$.
%{\color{red} [Include expressions for $\alpha_i$?]}
The couplings of each of the scalars in the model, $\lambda_i$, to the SM particles is scaled with respect to the SM scalar couplings $\lambda_{SM}$ by an element in $R$,

\begin{equation}
\lambda_i = R_{i1} \lambda_{SM}.
\label{eq:complex_coupling_scaled}
\end{equation}
The matrix $R$ also diagonalises the mass matrix $M^2$, resulting in the Higgs masses $m_{h_i}$,
\begin{equation}
\label{eq:complex_mass_diagonal}
R M^2 R^T = \mathrm{diag} \left( m_{h_1}, m_{h_2}, m_{h_3} \right).
\end{equation}
In the broken phase, the tree-level mass matrix $M$ takes the form,

\begin{equation}
\label{eq:complex_mass_matrix_broken}
M_{broken}^2 =
\begin{pmatrix}
\frac{\mu^2}{2} + \frac{\delta v_{s_1}^2}{4} + \frac{\delta v_{s_2}^2}{4} + \frac{3 \lambda v^2}{4} & \frac{\delta v v_{s_1}}{2} & \frac{\delta v v_{s_2}}{2} \\
\frac{\delta v v_{s_1}}{2} & b_{+} + \frac{3 d_2 v_{s_1}^2}{4} + \frac{d_2 v_{s_2}^2}{4} + \frac{\delta v^2}{4} & \frac{d_2 v_{s_1} v_{s_2}}{4} \\
\frac{\delta v v_{s_2}}{2} & \frac{d_2 v_{s_1} v_{s_2}}{4} & b_{-} + \frac{d_2 v_{s_1}^2}{4} + \frac{3 d_2 v_{s_2}^2}{4} + \frac{\delta v^2}{4}
\end{pmatrix},
\end{equation}
whereas in the DM phase where $v_{s_2} = 0$ the mass matrix becomes,

\begin{equation}
\label{eq:complex_mass_matrix_DM}
M_{DM}^2 =
\begin{pmatrix}
\frac{\mu^2}{2} + \frac{\delta v_{s_1}^2}{4} + \frac{3 \lambda v^2}{4} & \frac{\delta v v_{s_1}}{2} & 0 \\
\frac{\delta v v_{s_1}}{2} & b_{+} + \frac{3 d_2 v_{s_1}^2}{4} + \frac{\delta v^2}{4} & 0 \\
0 & 0 &  b_{-} + \frac{d_2 v_{s_1}^2}{4} + \frac{\delta v^2}{4}
\end{pmatrix}.
\end{equation}

\section{Renormalisation Group Equations of the Quartic Higgs Couplings}
\label{Appendix:RGEs}
The two-loop $\beta$-functions that describe the running of the SM Higgs couplings $\lambda$, the Higgs portal coupling $\delta$ and the complex singlet quartic coupling $d_2$, as calculated using SARAH \citep{Staub:2013tta}. Here $g_i$ are the SM gauge couplings and $Y_i$ are the Yukawa matrices.
{\allowdisplaybreaks  \begin{align}
\beta_{d_2}^{(1)} & =
2 \Big(10 d_{2}^{2}  + \delta^{2}\Big),\\
\beta_{d_2}^{(2)} & =
-\frac{4}{5} \Big(300 d_{2}^{3} +25 d_2 \delta^{2} -3 g_{1}^{2} \delta^{2} -15 g_{2}^{2} \delta^{2} +10 \delta^{3} +15 \delta^{2} \mbox{Tr}\Big({Y_d  Y_{d}^{\dagger}}\Big) +5 \delta^{2} \mbox{Tr}\Big({Y_e  Y_{e}^{\dagger}}\Big) \nonumber \\
 &+15 \delta^{2} \mbox{Tr}\Big({Y_u  Y_{u}^{\dagger}}\Big) \Big),\\
\beta_{\delta}^{(1)} & =
\frac{1}{10} \delta \Big(20 \mbox{Tr}\Big({Y_e  Y_{e}^{\dagger}}\Big)  + 40 \delta  -45 g_{2}^{2}  + 60 \lambda  + 60 \mbox{Tr}\Big({Y_d  Y_{d}^{\dagger}}\Big)  + 60 \mbox{Tr}\Big({Y_u  Y_{u}^{\dagger}}\Big)  + 80 d_2  -9 g_{1}^{2} \Big),\\
\beta_{\delta}^{(2)} & =
-40 d_{2}^{2} \delta +\frac{1671}{400} g_{1}^{4} \delta +\frac{9}{8} g_{1}^{2} g_{2}^{2} \delta -\frac{145}{16} g_{2}^{4} \delta -48 d_2 \delta^{2} +\frac{3}{5} g_{1}^{2} \delta^{2} +3 g_{2}^{2} \delta^{2} -11 \delta^{3} +\frac{36}{5} g_{1}^{2} \delta \lambda \nonumber \\
 &+36 g_{2}^{2} \delta \lambda -36 \delta^{2} \lambda -15 \delta \lambda^{2} +\frac{1}{4} \delta \Big(16 \Big(10 g_{3}^{2}  -3 \delta  -9 \lambda \Big) + 45 g_{2}^{2}  + 5 g_{1}^{2} \Big)\mbox{Tr}\Big({Y_d  Y_{d}^{\dagger}}\Big) \nonumber \\
 &+\frac{1}{4} \delta \Big(15 g_{1}^{2}  + 15 g_{2}^{2}  -16 \Big(3 \lambda  + \delta\Big)\Big)\mbox{Tr}\Big({Y_e  Y_{e}^{\dagger}}\Big) +\frac{17}{4} g_{1}^{2} \delta \mbox{Tr}\Big({Y_u  Y_{u}^{\dagger}}\Big) +\frac{45}{4} g_{2}^{2} \delta \mbox{Tr}\Big({Y_u  Y_{u}^{\dagger}}\Big) \nonumber \\
 &+40 g_{3}^{2} \delta \mbox{Tr}\Big({Y_u  Y_{u}^{\dagger}}\Big) -12 \delta^{2} \mbox{Tr}\Big({Y_u  Y_{u}^{\dagger}}\Big) -36 \delta \lambda \mbox{Tr}\Big({Y_u  Y_{u}^{\dagger}}\Big) -\frac{27}{2} \delta \mbox{Tr}\Big({Y_d  Y_{d}^{\dagger}  Y_d  Y_{d}^{\dagger}}\Big) \nonumber \\
 &-21 \delta \mbox{Tr}\Big({Y_d  Y_{u}^{\dagger}  Y_u  Y_{d}^{\dagger}}\Big) -\frac{9}{2} \delta \mbox{Tr}\Big({Y_e  Y_{e}^{\dagger}  Y_e  Y_{e}^{\dagger}}\Big) -\frac{27}{2} \delta \mbox{Tr}\Big({Y_u  Y_{u}^{\dagger}  Y_u  Y_{u}^{\dagger}}\Big), \\
\beta_{\lambda}^{(1)} & =
+\frac{27}{100} g_{1}^{4} +\frac{9}{10} g_{1}^{2} g_{2}^{2} +\frac{9}{4} g_{2}^{4} +2 \delta^{2} -\frac{9}{5} g_{1}^{2} \lambda -9 g_{2}^{2} \lambda +12 \lambda^{2} +12 \lambda \mbox{Tr}\Big({Y_d  Y_{d}^{\dagger}}\Big) +4 \lambda \mbox{Tr}\Big({Y_e  Y_{e}^{\dagger}}\Big) \nonumber \\
 &+12 \lambda \mbox{Tr}\Big({Y_u  Y_{u}^{\dagger}}\Big) -12 \mbox{Tr}\Big({Y_d  Y_{d}^{\dagger}  Y_d  Y_{d}^{\dagger}}\Big) -4 \mbox{Tr}\Big({Y_e  Y_{e}^{\dagger}  Y_e  Y_{e}^{\dagger}}\Big) -12 \mbox{Tr}\Big({Y_u  Y_{u}^{\dagger}  Y_u  Y_{u}^{\dagger}}\Big), \\
\beta_{\lambda}^{(2)} & =
-\frac{3411}{1000} g_{1}^{6} -\frac{1677}{200} g_{1}^{4} g_{2}^{2} -\frac{289}{40} g_{1}^{2} g_{2}^{4} +\frac{305}{8} g_{2}^{6} -8 \delta^{3} +\frac{1887}{200} g_{1}^{4} \lambda +\frac{117}{20} g_{1}^{2} g_{2}^{2} \lambda -\frac{73}{8} g_{2}^{4} \lambda -10 \delta^{2} \lambda \nonumber \\
 &+\frac{54}{5} g_{1}^{2} \lambda^{2} +54 g_{2}^{2} \lambda^{2} -78 \lambda^{3} \nonumber \\
 &+\frac{1}{10} \Big(225 g_{2}^{2} \lambda  -45 g_{2}^{4}  + 80 \Big(10 g_{3}^{2}  -9 \lambda \Big)\lambda  + 9 g_{1}^{4}  + g_{1}^{2} \Big(25 \lambda  + 54 g_{2}^{2} \Big)\Big)\mbox{Tr}\Big({Y_d  Y_{d}^{\dagger}}\Big) \nonumber \\
 &-\frac{3}{10} \Big(15 g_{1}^{4}  + 5 \Big(16 \lambda^{2}  -5 g_{2}^{2} \lambda  + g_{2}^{4}\Big) - g_{1}^{2} \Big(22 g_{2}^{2}  + 25 \lambda \Big)\Big)\mbox{Tr}\Big({Y_e  Y_{e}^{\dagger}}\Big) -\frac{171}{50} g_{1}^{4} \mbox{Tr}\Big({Y_u  Y_{u}^{\dagger}}\Big) \nonumber \\
 &+\frac{63}{5} g_{1}^{2} g_{2}^{2} \mbox{Tr}\Big({Y_u  Y_{u}^{\dagger}}\Big) -\frac{9}{2} g_{2}^{4} \mbox{Tr}\Big({Y_u  Y_{u}^{\dagger}}\Big) +\frac{17}{2} g_{1}^{2} \lambda \mbox{Tr}\Big({Y_u  Y_{u}^{\dagger}}\Big) +\frac{45}{2} g_{2}^{2} \lambda \mbox{Tr}\Big({Y_u  Y_{u}^{\dagger}}\Big) \nonumber \\
 &+80 g_{3}^{2} \lambda \mbox{Tr}\Big({Y_u  Y_{u}^{\dagger}}\Big) -72 \lambda^{2} \mbox{Tr}\Big({Y_u  Y_{u}^{\dagger}}\Big) +\frac{8}{5} g_{1}^{2} \mbox{Tr}\Big({Y_d  Y_{d}^{\dagger}  Y_d  Y_{d}^{\dagger}}\Big) -64 g_{3}^{2} \mbox{Tr}\Big({Y_d  Y_{d}^{\dagger}  Y_d  Y_{d}^{\dagger}}\Big) \nonumber \\
 &-3 \lambda \mbox{Tr}\Big({Y_d  Y_{d}^{\dagger}  Y_d  Y_{d}^{\dagger}}\Big) -42 \lambda \mbox{Tr}\Big({Y_d  Y_{u}^{\dagger}  Y_u  Y_{d}^{\dagger}}\Big) -\frac{24}{5} g_{1}^{2} \mbox{Tr}\Big({Y_e  Y_{e}^{\dagger}  Y_e  Y_{e}^{\dagger}}\Big) - \lambda \mbox{Tr}\Big({Y_e  Y_{e}^{\dagger}  Y_e  Y_{e}^{\dagger}}\Big) \nonumber \\
 &-\frac{16}{5} g_{1}^{2} \mbox{Tr}\Big({Y_u  Y_{u}^{\dagger}  Y_u  Y_{u}^{\dagger}}\Big) -64 g_{3}^{2} \mbox{Tr}\Big({Y_u  Y_{u}^{\dagger}  Y_u  Y_{u}^{\dagger}}\Big) -3 \lambda \mbox{Tr}\Big({Y_u  Y_{u}^{\dagger}  Y_u  Y_{u}^{\dagger}}\Big) +60 \mbox{Tr}\Big({Y_d  Y_{d}^{\dagger}  Y_d  Y_{d}^{\dagger}  Y_d  Y_{d}^{\dagger}}\Big) \nonumber \\
 &-24 \mbox{Tr}\Big({Y_d  Y_{d}^{\dagger}  Y_d  Y_{u}^{\dagger}  Y_u  Y_{d}^{\dagger}}\Big) +12 \mbox{Tr}\Big({Y_d  Y_{u}^{\dagger}  Y_u  Y_{d}^{\dagger}  Y_d  Y_{d}^{\dagger}}\Big) -12 \mbox{Tr}\Big({Y_d  Y_{u}^{\dagger}  Y_u  Y_{u}^{\dagger}  Y_u  Y_{d}^{\dagger}}\Big) \nonumber \\
 &+20 \mbox{Tr}\Big({Y_e  Y_{e}^{\dagger}  Y_e  Y_{e}^{\dagger}  Y_e  Y_{e}^{\dagger}}\Big) +60 \mbox{Tr}\Big({Y_u  Y_{u}^{\dagger}  Y_u  Y_{u}^{\dagger}  Y_u  Y_{u}^{\dagger}}\Big) .
\end{align}}

\end{appendices}

\bibliography{bibliography}{}

\providecommand{\href}[2]{#2}\begingroup\raggedright\begin{thebibliography}{10}

\bibitem{Aad:2012tfa}
{\bf ATLAS} Collaboration, G.~Aad et~al., {\it {Observation of a new particle
  in the search for the Standard Model Higgs boson with the ATLAS detector at
  the LHC}},  {\em Phys. Lett.} {\bf B716} (2012) 1--29,
  [\href{http://arxiv.org/abs/1207.7214}{{\tt arXiv:1207.7214}}].

\bibitem{Aad:2015zhl}
{\bf ATLAS, CMS} Collaboration, G.~Aad et~al., {\it {Combined Measurement of
  the Higgs Boson Mass in $pp$ Collisions at $\sqrt{s}=7$ and 8 TeV with the
  ATLAS and CMS Experiments}},  {\em Phys. Rev. Lett.} {\bf 114} (2015) 191803,
  [\href{http://arxiv.org/abs/1503.07589}{{\tt arXiv:1503.07589}}].

\bibitem{Buttazzo:2013uya}
D.~Buttazzo, G.~Degrassi, P.~P. Giardino, G.~F. Giudice, F.~Sala, A.~Salvio,
  and A.~Strumia, {\it {Investigating the near-criticality of the Higgs
  boson}},  {\em JHEP} {\bf 12} (2013) 089,
  [\href{http://arxiv.org/abs/1307.3536}{{\tt arXiv:1307.3536}}].

\bibitem{Craig:2013cxa}
N.~Craig, {\it {The State of Supersymmetry after Run I of the LHC}},
  \href{http://arxiv.org/abs/1309.0528}{{\tt arXiv:1309.0528}}.

\bibitem{Ross:2017kjc}
G.~G. Ross, K.~Schmidt-Hoberg, and F.~Staub, {\it {Revisiting fine-tuning in
  the MSSM}},  {\em JHEP} {\bf 03} (2017) 021,
  [\href{http://arxiv.org/abs/1701.03480}{{\tt arXiv:1701.03480}}].

\bibitem{Khachatryan:2014kca}
{\bf CMS} Collaboration, V.~Khachatryan et~al., {\it {Constraints on the
  spin-parity and anomalous HVV couplings of the Higgs boson in proton
  collisions at 7 and 8 TeV}},  {\em Phys. Rev.} {\bf D92} (2015), no.~1
  012004, [\href{http://arxiv.org/abs/1411.3441}{{\tt arXiv:1411.3441}}].

\bibitem{Khachatryan:2014jba}
{\bf CMS} Collaboration, V.~Khachatryan et~al., {\it {Precise determination of
  the mass of the Higgs boson and tests of compatibility of its couplings with
  the standard model predictions using proton collisions at 7 and 8 $\,\text
  {TeV}$}},  {\em Eur. Phys. J.} {\bf C75} (2015), no.~5 212,
  [\href{http://arxiv.org/abs/1412.8662}{{\tt arXiv:1412.8662}}].

\bibitem{Aad:2015gba}
{\bf ATLAS} Collaboration, G.~Aad et~al., {\it {Measurements of the Higgs boson
  production and decay rates and coupling strengths using pp collision data at
  $\sqrt{s}=7$ and 8 TeV in the ATLAS experiment}},  {\em Eur. Phys. J.} {\bf
  C76} (2016), no.~1 6, [\href{http://arxiv.org/abs/1507.04548}{{\tt
  arXiv:1507.04548}}].

\bibitem{Staub:2013tta}
F.~Staub, {\it {SARAH 4 : A tool for (not only SUSY) model builders}},  {\em
  Comput. Phys. Commun.} {\bf 185} (2014) 1773--1790,
  [\href{http://arxiv.org/abs/1309.7223}{{\tt arXiv:1309.7223}}].

\bibitem{Athron:2014yba}
P.~Athron, J.-h. Park, D.~Stockinger, and A.~Voigt, {\it {FlexibleSUSY - A
  spectrum generator generator for supersymmetric models}},  {\em Comput. Phys.
  Commun.} {\bf 190} (2015) 139--172,
  [\href{http://arxiv.org/abs/1406.2319}{{\tt arXiv:1406.2319}}].

\bibitem{Athron:2017fvs}
P.~Athron, M.~Bach, D.~Harries, T.~Kwasnitza, J.-h. Park, D.~Stöckinger,
  A.~Voigt, and J.~Ziebell, {\it {FlexibleSUSY 2.0: Extensions to investigate
  the phenomenology of SUSY and non-SUSY models}},
  \href{http://arxiv.org/abs/1710.03760}{{\tt arXiv:1710.03760}}.

\bibitem{Allanach:2001kg}
B.~C. Allanach, {\it {SOFTSUSY: a program for calculating supersymmetric
  spectra}},  {\em Comput. Phys. Commun.} {\bf 143} (2002) 305--331,
  [\href{http://arxiv.org/abs/hep-ph/0104145}{{\tt hep-ph/0104145}}].

\bibitem{Allanach:2013kza}
B.~C. Allanach, P.~Athron, L.~C. Tunstall, A.~Voigt, and A.~G. Williams, {\it
  {Next-to-Minimal SOFTSUSY}},  {\em Comput. Phys. Commun.} {\bf 185} (2014)
  2322--2339, [\href{http://arxiv.org/abs/1311.7659}{{\tt arXiv:1311.7659}}].

\bibitem{Degrassi:2012ry}
G.~Degrassi, S.~Di~Vita, J.~Elias-Miro, J.~R. Espinosa, G.~F. Giudice,
  G.~Isidori, and A.~Strumia, {\it {Higgs mass and vacuum stability in the
  Standard Model at NNLO}},  {\em JHEP} {\bf 08} (2012) 098,
  [\href{http://arxiv.org/abs/1205.6497}{{\tt arXiv:1205.6497}}].

\bibitem{Holthausen:2011aa}
M.~Holthausen, K.~S. Lim, and M.~Lindner, {\it {Planck scale Boundary
  Conditions and the Higgs Mass}},  {\em JHEP} {\bf 02} (2012) 037,
  [\href{http://arxiv.org/abs/1112.2415}{{\tt arXiv:1112.2415}}].

\bibitem{Iacobellis:2016eof}
G.~Iacobellis and I.~Masina, {\it {Stationary configurations of the Standard
  Model Higgs potential: electroweak stability and rising inflection point}},
  \href{http://arxiv.org/abs/1604.06046}{{\tt arXiv:1604.06046}}.

\bibitem{Haba:2013lga}
N.~Haba, K.~Kaneta, and R.~Takahashi, {\it {Planck scale boundary conditions in
  the standard model with singlet scalar dark matter}},  {\em JHEP} {\bf 04}
  (2014) 029, [\href{http://arxiv.org/abs/1312.2089}{{\tt arXiv:1312.2089}}].

\bibitem{Eichhorn:2014qka}
A.~Eichhorn and M.~M. Scherer, {\it {Planck scale, Higgs mass, and scalar dark
  matter}},  {\em Phys. Rev.} {\bf D90} (2014), no.~2 025023,
  [\href{http://arxiv.org/abs/1404.5962}{{\tt arXiv:1404.5962}}].

\bibitem{Khan:2014kba}
N.~Khan and S.~Rakshit, {\it {Study of electroweak vacuum metastability with a
  singlet scalar dark matter}},  {\em Phys. Rev.} {\bf D90} (2014), no.~11
  113008, [\href{http://arxiv.org/abs/1407.6015}{{\tt arXiv:1407.6015}}].

\bibitem{Helmboldt:2016mpi}
A.~J. Helmboldt, P.~Humbert, M.~Lindner, and J.~Smirnov, {\it {Minimal
  conformal extensions of the Higgs sector}},  {\em JHEP} {\bf 07} (2017) 113,
  [\href{http://arxiv.org/abs/1603.03603}{{\tt arXiv:1603.03603}}].

\bibitem{Darvishi:2016gvm}
N.~Darvishi and M.~Krawczyk, {\it {CP violation in the Standard Model with a
  complex singlet}},  \href{http://arxiv.org/abs/1603.00598}{{\tt
  arXiv:1603.00598}}.

\bibitem{Froggatt:1995rt}
C.~D. Froggatt and H.~B. Nielsen, {\it {Standard model criticality prediction:
  Top mass 173 +- 5-GeV and Higgs mass 135 +- 9-GeV}},  {\em Phys. Lett.} {\bf
  B368} (1996) 96--102, [\href{http://arxiv.org/abs/hep-ph/9511371}{{\tt
  hep-ph/9511371}}].

\bibitem{Froggatt:2006zc}
C.~D. Froggatt, L.~Laperashvili, R.~Nevzorov, H.~B. Nielsen, and M.~Sher, {\it
  {Implementation of the multiple point principle in the two-Higgs doublet
  model of type II}},  {\em Phys. Rev.} {\bf D73} (2006) 095005,
  [\href{http://arxiv.org/abs/hep-ph/0602054}{{\tt hep-ph/0602054}}].

\bibitem{Kawana:2014zxa}
K.~Kawana, {\it {Multiple Point Principle of the Standard Model with Scalar
  Singlet Dark Matter and Right Handed Neutrinos}},  {\em PTEP} {\bf 2015}
  (2015) 023B04, [\href{http://arxiv.org/abs/1411.2097}{{\tt
  arXiv:1411.2097}}].

\bibitem{Haba:2014sia}
N.~Haba, H.~Ishida, K.~Kaneta, and R.~Takahashi, {\it {Vanishing Higgs
  potential at the Planck scale in a singlet extension of the standard model}},
   {\em Phys. Rev.} {\bf D90} (2014) 036006,
  [\href{http://arxiv.org/abs/1406.0158}{{\tt arXiv:1406.0158}}].

\bibitem{Hamada:2014xka}
Y.~Hamada, H.~Kawai, and K.-y. Oda, {\it {Predictions on mass of Higgs portal
  scalar dark matter from Higgs inflation and flat potential}},  {\em JHEP}
  {\bf 07} (2014) 026, [\href{http://arxiv.org/abs/1404.6141}{{\tt
  arXiv:1404.6141}}].

\bibitem{Haba:2016gqx}
N.~Haba, H.~Ishida, N.~Okada, and Y.~Yamaguchi, {\it {Multiple-point principle
  with a scalar singlet extension of the Standard Model}},  {\em PTEP} {\bf
  2017} (2017), no.~1 013B03, [\href{http://arxiv.org/abs/1608.00087}{{\tt
  arXiv:1608.00087}}].

\bibitem{Nielsen:2017ows}
H.~F.~B. Nielsen, D.~L. Bennett, C.~R. Das, C.~D. Froggatt, and L.~V.
  Laperashvili, {\it {F(750), We Miss You as a Bound State of 6 Top and 6
  Antitop Quarks, Multiple Point Principle}},  {\em PoS} {\bf CORFU2016} (2017)
  050, [\href{http://arxiv.org/abs/1705.10749}{{\tt arXiv:1705.10749}}].

\bibitem{Shaposhnikov:2009pv}
M.~Shaposhnikov and C.~Wetterich, {\it {Asymptotic safety of gravity and the
  Higgs boson mass}},  {\em Phys. Lett.} {\bf B683} (2010) 196--200,
  [\href{http://arxiv.org/abs/0912.0208}{{\tt arXiv:0912.0208}}].

\bibitem{Wetterich:2011aa}
C.~Wetterich, {\it {Where to look for solving the gauge hierarchy problem?}},
  {\em Phys. Lett.} {\bf B718} (2012) 573--576,
  [\href{http://arxiv.org/abs/1112.2910}{{\tt arXiv:1112.2910}}].

\bibitem{Eichhorn:2017sok}
A.~Eichhorn, S.~Lippoldt, and V.~Skrinjar, {\it {Nonminimal hints for
  asymptotic safety}},  {\em Phys. Rev.} {\bf D97} (2018), no.~2 026002,
  [\href{http://arxiv.org/abs/1710.03005}{{\tt arXiv:1710.03005}}].

\bibitem{Eichhorn:2017lry}
A.~Eichhorn and F.~Versteegen, {\it {Upper bound on the Abelian gauge coupling
  from asymptotic safety}},  {\em JHEP} {\bf 01} (2018) 030,
  [\href{http://arxiv.org/abs/1709.07252}{{\tt arXiv:1709.07252}}].

\bibitem{Eichhorn:2017ylw}
A.~Eichhorn and A.~Held, {\it {Top mass from asymptotic safety}},  {\em Phys.
  Lett.} {\bf B777} (2018) 217--221,
  [\href{http://arxiv.org/abs/1707.01107}{{\tt arXiv:1707.01107}}].

\bibitem{Giudice:2014tma}
G.~F. Giudice, G.~Isidori, A.~Salvio, and A.~Strumia, {\it {Softened Gravity
  and the Extension of the Standard Model up to Infinite Energy}},  {\em JHEP}
  {\bf 02} (2015) 137, [\href{http://arxiv.org/abs/1412.2769}{{\tt
  arXiv:1412.2769}}].

\bibitem{Einhorn:2017jbs}
M.~B. Einhorn and D.~R.~T. Jones, {\it {Asymptotic freedom in certain $SO(N)$
  and $SU(N)$ models}},  {\em Phys. Rev.} {\bf D96} (2017), no.~5 055035,
  [\href{http://arxiv.org/abs/1705.00751}{{\tt arXiv:1705.00751}}].

\bibitem{Holdom:2014hla}
B.~Holdom, J.~Ren, and C.~Zhang, {\it {Stable Asymptotically Free Extensions
  (SAFEs) of the Standard Model}},  {\em JHEP} {\bf 03} (2015) 028,
  [\href{http://arxiv.org/abs/1412.5540}{{\tt arXiv:1412.5540}}].

\bibitem{Gies:2015lia}
H.~Gies and L.~Zambelli, {\it {Asymptotically free scaling solutions in
  non-Abelian Higgs models}},  {\em Phys. Rev.} {\bf D92} (2015), no.~2 025016,
  [\href{http://arxiv.org/abs/1502.05907}{{\tt arXiv:1502.05907}}].

\bibitem{Litim:2014uca}
D.~F. Litim and F.~Sannino, {\it {Asymptotic safety guaranteed}},  {\em JHEP}
  {\bf 12} (2014) 178, [\href{http://arxiv.org/abs/1406.2337}{{\tt
  arXiv:1406.2337}}].

\bibitem{Litim:2015iea}
D.~F. Litim, M.~Mojaza, and F.~Sannino, {\it {Vacuum stability of
  asymptotically safe gauge-Yukawa theories}},  {\em JHEP} {\bf 01} (2016) 081,
  [\href{http://arxiv.org/abs/1501.03061}{{\tt arXiv:1501.03061}}].

\bibitem{Bond:2016dvk}
A.~D. Bond and D.~F. Litim, {\it {Theorems for Asymptotic Safety of Gauge
  Theories}},  \href{http://arxiv.org/abs/1608.00519}{{\tt arXiv:1608.00519}}.

\bibitem{Bond:2017wut}
A.~D. Bond, G.~Hiller, K.~Kowalska, and D.~F. Litim, {\it {Directions for model
  building from asymptotic safety}},
  \href{http://arxiv.org/abs/1702.01727}{{\tt arXiv:1702.01727}}.

\bibitem{Sannino:2014lxa}
F.~Sannino and I.~M. Shoemaker, {\it {Asymptotically Safe Dark Matter}},  {\em
  Phys. Rev.} {\bf D92} (2015), no.~4 043518,
  [\href{http://arxiv.org/abs/1412.8034}{{\tt arXiv:1412.8034}}].

\bibitem{Bajc:2016efj}
B.~Bajc and F.~Sannino, {\it {Asymptotically Safe Grand Unification}},  {\em
  JHEP} {\bf 12} (2016) 141, [\href{http://arxiv.org/abs/1610.09681}{{\tt
  arXiv:1610.09681}}].

\bibitem{Pelaggi:2017wzr}
G.~M. Pelaggi, F.~Sannino, A.~Strumia, and E.~Vigiani, {\it {Naturalness of
  asymptotically safe Higgs}},  \href{http://arxiv.org/abs/1701.01453}{{\tt
  arXiv:1701.01453}}.

\bibitem{Pelaggi:2017abg}
G.~M. Pelaggi, A.~D. Plascencia, A.~Salvio, F.~Sannino, J.~Smirnov, and
  A.~Strumia, {\it {Asymptotically Safe Standard Model Extensions}},
  \href{http://arxiv.org/abs/1708.00437}{{\tt arXiv:1708.00437}}.

\bibitem{PDG:2015}
S.~Bethke, G.~Dissertori, and G.~Salam, {\it "particle data group review on
  quantum chromodynamics,
  \url{http://pdg.lbl.gov/2015/reviews/rpp2015-rev-qcd.pdf}"}, .

\bibitem{Ginsparg:1987ee}
P.~H. Ginsparg, {\it {Gauge and Gravitational Couplings in Four-Dimensional
  String Theories}},  {\em Phys. Lett.} {\bf B197} (1987) 139--143.

\bibitem{Witten:1996mz}
E.~Witten, {\it {Strong coupling expansion of Calabi-Yau compactification}},
  {\em Nucl. Phys.} {\bf B471} (1996) 135--158,
  [\href{http://arxiv.org/abs/hep-th/9602070}{{\tt hep-th/9602070}}].

\bibitem{Barger:2008jx}
V.~Barger, P.~Langacker, M.~McCaskey, M.~Ramsey-Musolf, and G.~Shaughnessy,
  {\it {Complex Singlet Extension of the Standard Model}},  {\em Phys. Rev.}
  {\bf D79} (2009) 015018, [\href{http://arxiv.org/abs/0811.0393}{{\tt
  arXiv:0811.0393}}].

\bibitem{Barger:2010yn}
V.~Barger, M.~McCaskey, and G.~Shaughnessy, {\it {Complex Scalar Dark Matter
  vis-\`{a}-vis CoGeNT, DAMA/LIBRA and XENON100}},  {\em Phys. Rev.} {\bf D82}
  (2010) 035019, [\href{http://arxiv.org/abs/1005.3328}{{\tt
  arXiv:1005.3328}}].

\bibitem{Gonderinger:2012rd}
M.~Gonderinger, H.~Lim, and M.~J. Ramsey-Musolf, {\it {Complex Scalar Singlet
  Dark Matter: Vacuum Stability and Phenomenology}},  {\em Phys. Rev.} {\bf
  D86} (2012) 043511, [\href{http://arxiv.org/abs/1202.1316}{{\tt
  arXiv:1202.1316}}].

\bibitem{Costa:2014qga}
R.~Costa, A.~P. Morais, M.~O.~P. Sampaio, and R.~Santos, {\it {Two-loop
  stability of a complex singlet extended Standard Model}},  {\em Phys. Rev.}
  {\bf D92} (2015) 025024, [\href{http://arxiv.org/abs/1411.4048}{{\tt
  arXiv:1411.4048}}].

\bibitem{Coimbra:2013qq}
R.~Coimbra, M.~O.~P. Sampaio, and R.~Santos, {\it {ScannerS: Constraining the
  phase diagram of a complex scalar singlet at the LHC}},  {\em Eur. Phys. J.}
  {\bf C73} (2013) 2428, [\href{http://arxiv.org/abs/1301.2599}{{\tt
  arXiv:1301.2599}}].

\bibitem{Robens:2015gla}
T.~Robens and T.~Stefaniak, {\it {Status of the Higgs Singlet Extension of the
  Standard Model after LHC Run 1}},  {\em Eur. Phys. J.} {\bf C75} (2015) 104,
  [\href{http://arxiv.org/abs/1501.02234}{{\tt arXiv:1501.02234}}].

\bibitem{Muhlleitner:2017dkd}
M.~Muhlleitner, M.~O.~P. Sampaio, R.~Santos, and J.~Wittbrodt, {\it
  {Phenomenological Comparison of Models with Extended Higgs Sectors}},
  \href{http://arxiv.org/abs/1703.07750}{{\tt arXiv:1703.07750}}.

\bibitem{Camargo-Molina:2013qva}
J.~E. Camargo-Molina, B.~O'Leary, W.~Porod, and F.~Staub, {\it
  {$\mathbf{Vevacious}$: A Tool For Finding The Global Minima Of One-Loop
  Effective Potentials With Many Scalars}},  {\em Eur. Phys. J.} {\bf C73}
  (2013), no.~10 2588, [\href{http://arxiv.org/abs/1307.1477}{{\tt
  arXiv:1307.1477}}].

\bibitem{Bechtle:2013wla}
P.~Bechtle, O.~Brein, S.~Heinemeyer, O.~Stal, T.~Stefaniak, G.~Weiglein, and
  K.~E. Williams, {\it {$\mathsf{HiggsBounds}-4$: Improved Tests of Extended
  Higgs Sectors against Exclusion Bounds from LEP, the Tevatron and the LHC}},
  {\em Eur. Phys. J.} {\bf C74} (2014), no.~3 2693,
  [\href{http://arxiv.org/abs/1311.0055}{{\tt arXiv:1311.0055}}].

\bibitem{Bechtle:2013xfa}
P.~Bechtle, S.~Heinemeyer, O.~Stal, T.~Stefaniak, and G.~Weiglein, {\it
  {$HiggsSignals$: Confronting arbitrary Higgs sectors with measurements at the
  Tevatron and the LHC}},  {\em Eur. Phys. J.} {\bf C74} (2014), no.~2 2711,
  [\href{http://arxiv.org/abs/1305.1933}{{\tt arXiv:1305.1933}}].

\bibitem{Costa:2015llh}
R.~Costa, M.~Mühlleitner, M.~O.~P. Sampaio, and R.~Santos, {\it {Singlet
  Extensions of the Standard Model at LHC Run 2: Benchmarks and Comparison with
  the NMSSM}},  {\em JHEP} {\bf 06} (2016) 034,
  [\href{http://arxiv.org/abs/1512.05355}{{\tt arXiv:1512.05355}}].

\bibitem{Djouadi:1997yw}
A.~Djouadi, J.~Kalinowski, and M.~Spira, {\it {HDECAY: A Program for Higgs
  boson decays in the standard model and its supersymmetric extension}},  {\em
  Comput. Phys. Commun.} {\bf 108} (1998) 56--74,
  [\href{http://arxiv.org/abs/hep-ph/9704448}{{\tt hep-ph/9704448}}].

\bibitem{Butterworth:2010ym}
J.~M. Butterworth et~al., {\it {THE TOOLS AND MONTE CARLO WORKING GROUP Summary
  Report from the Les Houches 2009 Workshop on TeV Colliders}},  in {\em
  {Physics at TeV colliders. Proceedings, 6th Workshop, dedicated to Thomas
  Binoth, Les Houches, France, June 8-26, 2009}}, 2010.
\newblock \href{http://arxiv.org/abs/1003.1643}{{\tt arXiv:1003.1643}}.

\bibitem{Dawson:2017jja}
S.~Dawson and M.~Sullivan, {\it {Enhanced di-Higgs boson production in the
  complex Higgs singlet model}},  {\em Phys. Rev.} {\bf D97} (2018), no.~1
  015022, [\href{http://arxiv.org/abs/1711.06683}{{\tt arXiv:1711.06683}}].

\bibitem{Belanger:2014vza}
G.~Bélanger, F.~Boudjema, A.~Pukhov, and A.~Semenov, {\it {micrOMEGAs4.1: two
  dark matter candidates}},  {\em Comput. Phys. Commun.} {\bf 192} (2015)
  322--329, [\href{http://arxiv.org/abs/1407.6129}{{\tt arXiv:1407.6129}}].

\bibitem{Hinshaw:2012aka}
{\bf WMAP} Collaboration, G.~Hinshaw et~al., {\it {Nine-Year Wilkinson
  Microwave Anisotropy Probe (WMAP) Observations: Cosmological Parameter
  Results}},  {\em Astrophys. J. Suppl.} {\bf 208} (2013) 19,
  [\href{http://arxiv.org/abs/1212.5226}{{\tt arXiv:1212.5226}}].

\bibitem{Ade:2015xua}
{\bf Planck} Collaboration, P.~A.~R. Ade et~al., {\it {Planck 2015 results.
  XIII. Cosmological parameters}},  {\em Astron. Astrophys.} {\bf 594} (2016)
  A13, [\href{http://arxiv.org/abs/1502.01589}{{\tt arXiv:1502.01589}}].

\bibitem{Akerib:2016vxi}
{\bf LUX} Collaboration, D.~S. Akerib et~al., {\it {Results from a search for
  dark matter in the complete LUX exposure}},  {\em Phys. Rev. Lett.} {\bf 118}
  (2017), no.~2 021303, [\href{http://arxiv.org/abs/1608.07648}{{\tt
  arXiv:1608.07648}}].

\bibitem{Ellis:2000ds}
J.~R. Ellis, A.~Ferstl, and K.~A. Olive, {\it {Reevaluation of the elastic
  scattering of supersymmetric dark matter}},  {\em Phys. Lett.} {\bf B481}
  (2000) 304--314, [\href{http://arxiv.org/abs/hep-ph/0001005}{{\tt
  hep-ph/0001005}}].

\bibitem{Baker:2015qna}
M.~J. Baker et~al., {\it {The Coannihilation Codex}},  {\em JHEP} {\bf 12}
  (2015) 120, [\href{http://arxiv.org/abs/1510.03434}{{\tt arXiv:1510.03434}}].

\bibitem{Ghorbani:2014gka}
K.~Ghorbani and H.~Ghorbani, {\it {Scalar split WIMPs in future direct
  detection experiments}},  {\em Phys. Rev.} {\bf D93} (2016), no.~5 055012,
  [\href{http://arxiv.org/abs/1501.00206}{{\tt arXiv:1501.00206}}].

\bibitem{Chang:2017ynj}
W.-F. Chang, T.~Modak, and J.~N. Ng, {\it {Signal for a light singlet scalar at
  the LHC}},  \href{http://arxiv.org/abs/1711.05722}{{\tt arXiv:1711.05722}}.

\end{thebibliography}\endgroup
\bibliographystyle{JHEP}

\end{document}